# High and dry: billion-year trends in the aridity of river-forming climates on Mars

Edwin S. Kite[1], Axel Noblet[2,3]

1. University of Chicago, Department of the Geophysical Sciences, Chicago, IL 60637, USA.
2. Nantes Université, LPG-CNRS-UMR6112, Nantes, France.
3. University of Western Ontario, Inst. Earth & Space Exploration, Canada.

Corresponding author: Edwin Kite (kite@uchicago.edu)

**Key Points:**

- Aridity on Mars increased over time, but intermittently wetter climates persisted in lowlands
- Consistent with a change in Mars' greenhouse effect that left highlands too cold for liquid water except during a brief melt season
- Data are consistent with switch, of unknown cause, in dependence of aridity index on elevation: high-and-wet early on, high-and-dry later

**Abstract**

Mars' wet-to-dry transition is a major environmental catastrophe, yet the spatial pattern, tempo, and cause of drying are poorly constrained. We built a globally-distributed database of constraints on Mars late-stage paleolake size relative to catchment area (aridity index), and found evidence for climate zonation as Mars was drying out. Aridity increased over time in southern midlatitude highlands, where lakes became proportionally as small as in modern Nevada. Meanwhile, intermittently wetter climates persisted in equatorial and northern-midlatitude lowlands. This is consistent with a change in Mars' greenhouse effect that left highlands too cold for liquid water except during a brief melt season, or alternatively with a fall in Mars' groundwater table. The data are consistent with a switch of unknown cause in the dependence of aridity index on elevation, from high-and-wet early on, to high-and-dry later. These results sharpen our view of Mars' climate as surface conditions became increasingly stressing for life.

**Plain Language Summary**

Mars' surface was habitable in the past but is sterile today. Mars had multiple lake-forming eras as the planet dried out, but so far, there has been no globally distributed survey of the size of late-stage lakes, and the evaporation/precipitation ratio (aridity index) of the climates that formed them. This is key input/test data for models of Mars' past climate and climate evolution. We built a globally-distributed database of aridity index constraints for late-stage river-forming climates on Mars. On average, late-stage lake-forming climates had a higher aridity than early-stage river-forming climates. Drying-out was spatially heterogenous, with a "high-and-dry" pattern. This apparently contrasts with a "high-and-wet" pattern seen for early-stage river-forming climates. The reasons for this apparent switch are unknown.

# 1 Introduction

Today Mars is a cold desert, but billions of years ago Mars had rivers and lakes. Early on, during the Late Noachian / Early Hesperian (~3.6 Ga), water supply to crater lakes was large enough relative to evaporation that – at least intermittently – liquid water overspilled to carve canyons (Fassett & Head 2008). Later, runoff continued intermittently for ≳1 Gyr (e.g., Holo et al. 2021, Kite 2019, Grant & Wilson 2011), forming deltas and alluvial fans (e.g. Grant & Wilson 2011, Salese et al. 2019) that were probably precipitation-fed (Kite 2019), but these features were patchy (Wilson et al. 2021), with relatively few aqueous minerals visible from orbit (Pan et al. 2021), and lake overspills were less frequent (Goudge et al. 2016). At least some of the delta materials to be returned to Earth from Jezero crater likely date from the later era (e.g. Mangold et al. 2020, Salese et al. 2020). The data suggest a shift over time to wet events that were more short-lived, and/or to more arid climates. During this period, Mars was losing both $CO_2$ and $H_2O$, and the rate of asteroid impacts had declined to near-modern levels, yet volcanism and chaotic large-amplitude obliquity change continued (Haberle et al. 2017). Understanding the cause of changing lake levels is key to understanding the habitable-to-uninhabitable transition of Mars' surface environment, but the change itself is, as yet, poorly quantified.

To understand Mars' wet-to-dry transition, we need to know trends over time in mean aridity, and the spatial distribution of aridity. Past aridity (specifically, aridity index, the ratio of potential evaporation to precipitation) can be constrained using paleolake size. The topographic catchment area feeding into the lake divided by paleolake size (the hydrologic X-ratio, $X_H$) is, in hydrological steady state, approximately equal to the climatic aridity index (AI). This is because water from the topographic catchment is routed into a small area (the lake), and evaporation from the basin is reduced in proportion to the smallness of the lake (Matsubara & Howard 2009). Following (e.g.) Stucky de Quay et al. (2020), we assume that all meltwater/rainwater is routed to the lake, infiltration is minor, the lake level is in hydrologic steady state, and runoff production on the lake itself is small (so that $X_H$ = (topographic catchment area not including lake area) / lake area). Aridity index constrains paleoclimate models (Turbet & Forget 2021), and is a window into the evolution of ancient climate on Mars. Earlier-stage river-forming climates had $X_H$ = 3-7 (Fassett & Head 2008, Matsubara et al. 2011, Stucky de Quay et al. 2020), perhaps less (Matsubara et al. 2013). However, perhaps surprisingly, few estimates exist for how aridity index changed over time on Mars (e.g. Horvath & Andrews-Hanna 2017), and there has been no previous survey of $X_H$ for late-stage rivers/lakes.

In this study, we surveyed the interiors of all large, young craters (*n* = 212 areas) at latitudes between 40°N and 40°S mapped as Late Hesperian or Amazonian impacts (from Tanaka et al. 2014). The purpose of the latitude cut was to minimize overprinting by ice-associated processes. We also surveyed 7 additional craters that have relatively well-preserved rims, denoting relative youth, and defining closed basins. Well-preserved rims correspond to a closed catchment in most cases. The water-worn landforms within these craters formed during the Late Hesperian and Amazonian - extending more than 1 Gyr after the valley networks (e.g., Grant & Wilson 2011). Many workers have argued that the wet events were probably intermittent (e.g. Kite 2019, and references therein), with no rivers flowing for most of the time. Our results constrain conditions shortly before the last drying-out of low-latitude rivers on Mars. We then compared to work on early-stage rivers to find aridity trends over time.

## 2 Materials and Methods

To search for paleohydrologic proxies, we used Context Camera (CTX) data (Malin et al. 2007, Dickson et al. 2018), supplemented in a few places by High Resolution Imaging Science Experiment (HiRISE) (McEwen et al. 2007) images. For topography, we used Mars Orbiter Laser Altimeter (MOLA) Precision Experimental Data Records (PEDRs) (Smith et al. 2001), and in some places we used CTX /HiRISE DTMs (Mayer & Kite 2016).

To constrain $X_H$, we need estimates of paleolake area, $A$, and drainage area, $D$ ($X_H=(D-A)/A=D/A-1$; Matsubara et al. 2011) (Fig. 1). Here the "-1" corresponds to the assumption that no runoff is produced on the lake itself. To get paleolake area, thanks to the high quality of Mars topographic data (Smith et al. 2001) together with only minor post-lacustrine modification, it is usually enough to know paleolake water level. However, water level changed over time and most geologic proxies for past water level on Mars are indirect, so these estimates are not precise and may correspond to only the maximum (highest) lake levels. From the water level, the contour-enclosed area corresponds to past lake area. Water level constraints include flat crater-bottom deposits interpreted as playa/lake deposits, intra-basin spillways, and delta break-in-slope elevations. Upper-bound constraints come from the lowest (terminal) elevations of subaerial fans and channels, as these cannot form below lake level. For alluvial fan toes/channel termini, we used low-point elevations to draw an enclosing contour. The areas enclosed by these contours are upper limits on $A$. Many fans on Mars formed over long timescales (Kite et al. 2017), but channels can be carved rapidly (e.g., Whipple et al. 2000). Therefore, the $X_H$ obtained from setting lake elevation equal to a channel-stop elevation (requiring that the lake level did not exceed the channel-stop elevation for at least as long as it took to carve the channel) constrains lake level over a shorter timescale than the constraint obtained from a fan terminus elevation. (In figures, we mark the shorter-timescale channel-stop lower limits on $X_H$ with red open triangles, and the longer-timescale fan toe lower limits on $X_H$ with red filled triangles.) Wind erosion reduces lake deposit extent relative to original extent; we use lake deposit area as a lower limit on $A$. Internal spillways also provide lower limits on A. The slope-break elevation of deltas provides a best estimate of past lake level. For the deltas we analyzed layer-orientation data was not available, so our assessment was based only on geomorphic expression (modern topography), which is less definitive than stratigraphic methods (e.g. Tebolt & Goudge 2022). Our approach treats the present-day topographic relationships between lake deposit outcrops and alluvial fan deposit outcrops as being representative of the topographic relationships between deposits when the rivers were flowing. Infrequently, we observe flat crater-bottom deposits topographically above fan toes (e.g. at Luba crater), presumably due to differential wind erosion. At Peridier, channels extend topographically below the flat crater-bottom deposits that we interpret as lake deposits, perhaps corresponding to a later wet episode. Small channels were neglected. We recorded only the constraints that (within a given drainage area) were the most hydrologically constraining.

Drainage areas were taken to be the entire area of the host crater, except when internal drainage relations (ridges, spillways) showed a smaller contributing area. Some crater interiors contained multiple drainage areas, due to internal drainage divides, which were accounted for separately. We assume that all topographic drainage area (and not just the area upstream of observed channels) contributes water to the lake. If water was sourced by patchy snowmelt, then more runoff production per unit area would be needed.

Each proxy type has been described previously in detailed studies (e.g., Moore & Howard 2005, Palucis et al. 2016). For example, flat-lying sediments interpreted as lake deposits are described by Grant et al. (2008) and Morgan et al. (2014). As far as we know, this is the first global survey for flat crater-bottom deposits (FCBDs) that we interpret as lake deposits, so we provide more information on this proxy type below. Figs. S1-S4 show many-km-wide crater-bottom deposits that have an elevation range of only a few meters. Flat crater-bottom deposits frequently have slopes of 1/1000, 10-20× flatter than nearby alluvial fans. Flat crater-bottom deposits lack channels, show pits and grooves due to wind erosion, are usually found downslope from alluvial fans, and are typically bounded by outward-facing scarps. Internal layering and susceptibility to wind erosion suggest that these are indurated sedimentary deposits. Extreme flatness and location at the bottom of a crater indicate that aggradation was controlled by an equipotential, most likely liquid water. In some cases (Fig. S2a), sinuous ridges that we interpret as capped by fluvial deposits connect to FCBDs (Davis et al. 2019), strengthening confidence in the lake-deposit interpretation of those FCBDs. HiRISE DTMs (Figs. S2-S4, S7) confirm these impressions and adds detail on the gentle tilts of internal layers (typically ≲1°, consistent with flat once tracing and DTM errors are taken into account). These dips are significantly lower than those typical for Mars sediments interpreted as topography-draping air fall deposits (Annex & Lewis 2020). Sediments may have been transported into the lakes by either wind or water. Internal-layer conformity with deposit tops proves that flatness of deposit tops is not a chance of erosion but rather a trace of depositional process. Layers are expressed due to contrasts in erosional resistance, which in turn might relate to changes in grainsize or composition. Although no spectral confirmation of aqueous minerals is available in most cases, and therefore it remains conceivable that some of these FCBDs might be impact melt, we interpret these as lake deposits. (Unlike impact melts, catalogued FCBDs are flatter, and lack flow texture, arcuate ridges, and crumple ridges). When topographic data were lacking, we marked the deposit as a "candidate" FCBD. It is likely that some lake deposits are not interpretable from orbiter data (false negatives). The Murray mudstones at Gale, interpreted as lake deposits by Grotzinger et al. (2015), would not be counted this way.

## 3 Results of survey: latitude and elevation trends

Most craters show evidence for past liquid water ($n = 118$ basins) (Fig. 1). Presumably some craters lack evidence for liquid water because they postdate Mars' drying-out. Past lake size is constrained by the extent of flat crater-bottom deposits (FCBD) interpreted as lake/playa deposits (e.g., Morgan et al. 2014) ($n=87$ including 30 candidates; Figs. S1-S4), as well as the elevations of features such as alluvial fan termini, channel termini, internal spillways, candidate shorelines, and the break-in-slope elevation of scarp-fronted deposits interpreted as deltas ($n = 135$) (Fig. 1).

Within craters, fans were built by flows from crater sidewalls, suggesting precipitation runoff was responsible for fluvial sediment transport (Lamb et al. 2006). Lake levels could have been maintained by water conveyed from crater sidewalls (by surface runoff or shallow groundwater flow) or alternatively, by deep-upwelling groundwater sourced from (e.g.) rain/melt recharge at ~$10^2$-$10^3$-km scales (Salese et al. 2019, Horvath & Andrews-Hanna 2017). We do not think that the lakes were flooded by catastrophic release of groundwater (Wang et al. 2005), because channels run up to ridgelines and because only some places saw water. Fig. 2a shows the nonuniform distribution of craters with evidence for liquid water. Because we only survey craters that formed relatively recently (Tanaka et al. 2014), our survey provides strong evidence that relatively recent rivers/lakes were more frequent at off-equatorial latitudes (Fig. 2b) (Wilson et al.

2021). Lower-lying craters more frequently show evidence for past rivers/lakes (Fig. S6b) (Kite et al. 2022).

Figs. 3-4 sum up survey results. Typical $X_H$ was 7-38 (median fan terminus constraint to median lake deposit constraint), corresponding to arid-to-hyperarid conditions, with deltas and overspill channels recording semi-arid conditions (median $X_H = 4$). This is more arid than that reported for early lakes (by e.g. Matsubara et al. 2011), $X_{H,ancient} = 5\pm2$, and is similar to that of modern Western Nevada ($X_H = 19.7$ according to Matsubara et al. 2011). In the southern midlatitudes (high ground), late-stage $X_H < 10$ is less common south of 10°S (Fig. 4) – in other words, there is very little evidence for conditions moister than modern Nevada. By contrast, in the northern midlatitudes and at the equator (lower ground), the break-in-slope elevations of scarp-fronted deposits interpreted as deltas (filled blue diamonds in Fig. 4) indicate relatively big lakes. The difference between hemispheres gives a $p$-value of 0.002, and $X_H < 10$ is found mostly at <-1500 m elevation ($p = 0.01$). In summary, at high elevations, the central aridity estimate is more arid than modern Nevada. At lower elevations, some locations were less arid than modern Nevada (at least intermittently), similar to aridity estimates for the earlier-stage river era.

**4. Aridity change with time: high and dry.**

Late-stage rivers were apparently more spatially patchy than early-stage rivers. Where runoff did occur, we find (on average) a record of more arid climates. For the early-stage lakes (~3.6 Ga), Matsubara et al. (2011) report $X_{H,ancient} = 5\pm2$. This is comparable to the $X_H$ for the U.S. Great Basin wet period ~20 Kya ($X_H \approx 3.5$ according to Matsubara et al. 2011). These data indicate a climate trend, to more arid climate (our preferred explanation), or alternatively toward briefer wet events (disfavored by our data, Supplementary Information), on global average. Southern midlatitudes show a shift over time towards smaller lakes, but intermittent wet climates (semiarid to arid) persisted near the equator and in the northern midlatitudes. Aridity increased greatly at high elevations but only slightly at low elevations (Fig. 5). Similar $X_H$ at low elevations during the Noachian through Amazonian is consistent with the paucity of late-stage low-latitude lake overspills (Goudge et al. 2016). The high-relief rims of the young craters surveyed in this study make overspill more difficult for a given $X_H$ than for the more muted rims of ancient craters.

Intriguingly, Stucky de Quay et al. (2020)'s analysis of early-stage lake overspills shows (their Fig. 4) the strongest requirements for humid conditions at high elevation (Fig. S9). Stucky de Quay et al. (2020)'s 96-paleolake dataset consists of "hydrological systems in which the morphologies indicated precipitation as a main water source, either as rain or snow […] open- and closed-basin lakes fed by dendritic valley networks with a main trunk having a Strahler order of ≥3" (the less selective dataset of Fassett & Head (2008) shows only a weak trend toward more-humid conditions at high elevation). Stucky de Quay et al.'s (2020) result is the opposite of our late-stage result, suggesting a reversal of the elevation dependence of $X$-ratio as Mars evolved.

**5 Discussion and conclusions**

The late-stage trend to higher $X_H$ at high elevation (Fig. 3) can be explained by a change in the strength of greenhouse warming over time, such that highlands became almost always too cold for liquid water (Kite et al. 2022). In this scenario, higher-elevation lakes would have less meltwater runoff. Alternatively, in a very-warm-climate scenario (too warm for ground ice), if infiltration became an important water sink for lakes, then high lakes would lose water while water would upwell at low elevations. Thus, a decline in Mars' groundwater table (e.g., Andrews-Hanna & Lewis 2011, Salese et al. 2019, Jakosky et al. 2021) might also explain the Fig. 3 trends. A third possibility is that water vapor was sourced from evaporation at very low elevations, and that highlands were most horizontally distant from the water source and therefore water-starved (Turbet & Forget 2019). We favor the snow/ice-melt explanation because snow/ice-melt can be patchy and the late-stage erosion is patchy. Snow/ice melt is consistent with evidence for equatorial thermokarst (Warner et al. 2010).

In the future, measurements of the grain size of clasts moved by late-stage rivers might be decisive in distinguishing between the drying-while-warm versus drying-while-cool scenarios for Mars. At Gale crater, the rover *Curiosity* has encountered proxies for aridity such as mudcracks (Stein et al. 2018). Our data include late-stage Gale lakes (Palucis et al. 2016), conceivably preserved at Gediz Vallis ridge. Thus *Curiosity*'s traverse may allow ground-truthing of the scenario in Fig. 5.

Our results reinforce the interpretation (e.g., Irwin et al. 2015) that a hydrologic cycle fueled late-stage rivers. River/lake sediments found within large "host" impact craters often encapsulate smaller impact craters, which today appear partially exhumed (Table S1). In order for these smaller craters to accumulate, a time interval of ≥0.2 Gyr between the formation of the "host" craters and the end of river activity is required. This disproves the hypothesis (Mangold et al. 2012) that these rivers were triggered by the energy of the impact that formed the "host" crater. This is because ≥0.2 Gyr is too long for the energy of the "host"-crater-forming impact to contribute to fan formation (Kite et al. 2017; Table S2). The distribution of $X_H$ with crater size suggests wet events lasting at least decades (Supplementary Material).

The northern hemisphere $X_H$ permits a late-stage ocean, consistent with some models (e.g., Di Achille & Hynek 2010, Schmidt et al. 2022). However, the case for a Mars ocean remains equivocal: delta locations suggest deltas drained into large lakes, not an ocean (e.g. Rivera-Hernández and Palucis 2019).

Mars has many relatively-young exit-breach craters or "pollywogs" (Wilson et al. 2016). Almost all are omitted from of our study, because of their small size and location poleward of 40°. Pollywog overspill (Warren et al. 2020) would suggest $X_H < 1$, very different from the aridity of the low/mid latitudes, or, alternatively, groundwater release. It also remains to be determined whether late-stage lake overspills in Valles Marineris (Warner et al. 2013) match the within-crater $X_H$ pattern (Fig. 4).

A limitation of our study is that we do not distinguish between playas and perennial lakes. However, if smaller lakes dried up seasonally, then that would make the annual-average climate even more arid than reported here, and accentuate the "high-and-dry" pattern, so this limitation is not severe.

In summary, a globally distributed survey of paleohydrologic proxies for late-stage river-forming climates (Figs. 1-2), when compared to previous work on early-stage river-forming climates (Fig. 3), indicates a climate trend, to more arid climate (our preferred explanation), or, alternatively, toward briefer wet events (disfavored by data; Supplementary Information). Southern midlatitudes show a shift toward smaller lakes over time, but intermittent wetter climates persisted near the equator and in the northern midlatitudes. These results sharpen our view of Mars' wet-to-dry transition, but overall, it is surprising that this major environmental catastrophe remains so poorly understood. The challenge to models of Mars' climate (e.g., Turbet & Forget 2021, Kite et al. 2021) and climate evolution (e.g., Ramirez & Craddock 2018, Wordsworth et al. 2021) is now to explain these data.

**Acknowledgments**
We thank Lu Pan, Michael Mischna, Susan Conway, Alexandra Warren, and HiWish. We affirm that there are no financial (or other) conflicts of interest. Funding: NASA(80NSSC20K0144+80NSSC18K1476).

**Availability statement**
Consistent with https://www.agu.org/Publish-with-AGU/Publish/Author-Resources/Data-and-Software-for-Authors, data are shared in Supplementary Tables accompanying this submitted manuscript. The files constituting the database, the DTMs generated for this study, and other supporting files, are shared using Open Science Framework (osf.io) at https://osf.io/bue4m/ (DOI 10.17605/OSF.IO/BUE4M) (Kite & Noblet, 2022).

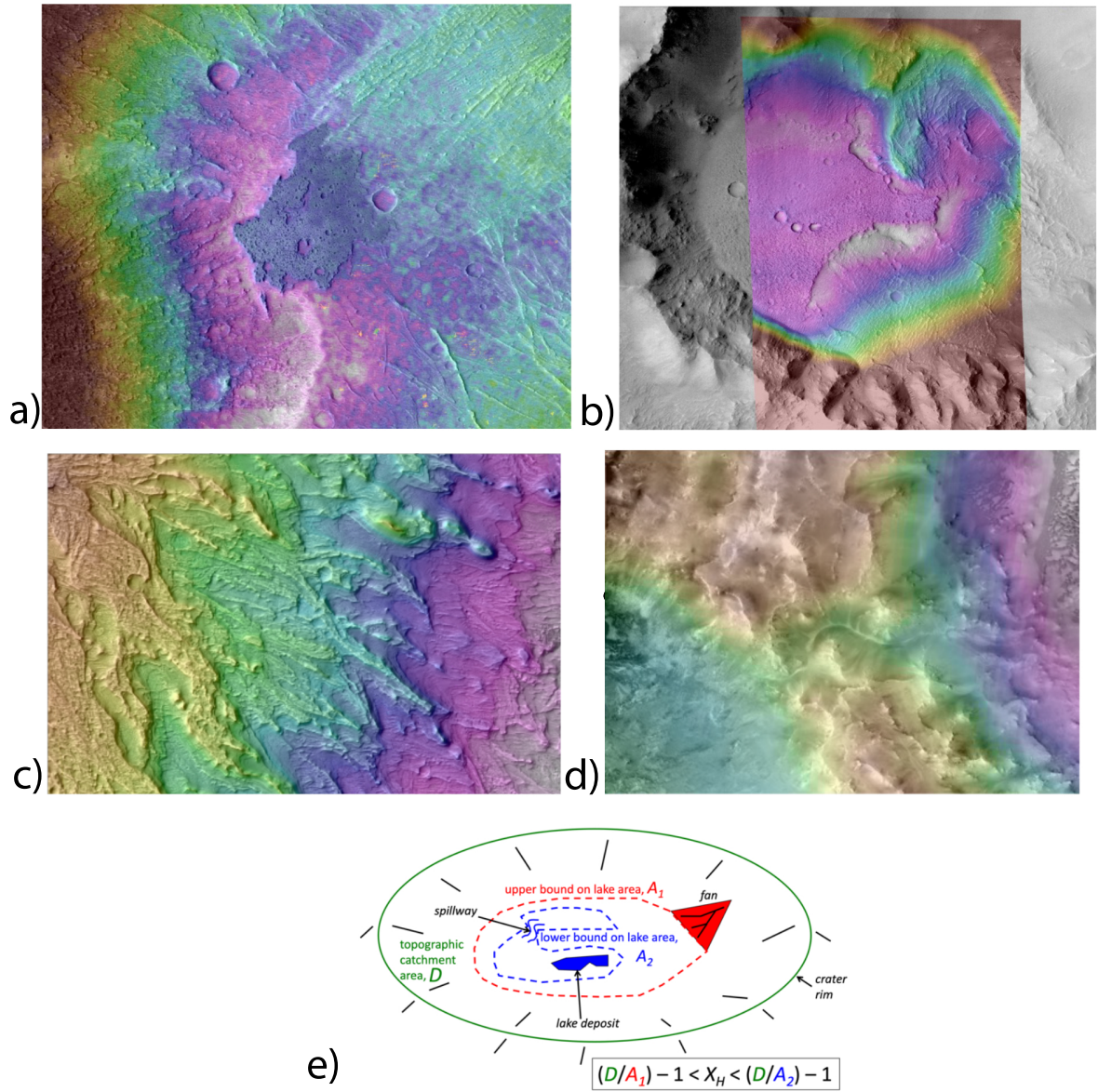

**Fig. 1.** Examples of paleohydrologic proxies (for details, see Figs. S1-S4). **(a)** Flat crater-bottom deposit (arrows) interpreted as a lake/playa deposit. Image is ~10 km across. Elevation range 150m. F10_039889_1567_XN_23S286W stereopair. 23°S 74°E. **(b)** Additional flat crater-bottom deposit (arrows) interpreted as a lake/playa deposit. Colored HiRISE DTM (ESP_065414_1495/ESP_065480_1495 stereopair) is 5.2 km across. Red-to-white elevation range is 200m. The S rim of the impact crater has numerous erosional alcoves, linked by a depositional ramp to the flat crater-bottom deposit. The depositional ramp is topped by sinuous ridges, one of which feeds into the flat crater-bottom deposit. **(c)** Alluvial fan deposit (image is 5.5 km across, PSP_007688_1575/ PSP_008545_1575 stereopair, elevation range 200m). **(d)** Spillway (arrow) (image is 29.7 km across, J15_050464_1996_XI_19N284W, elevation range 2260 m). **(e)** Cartoon showing use of paleohydrologic proxy data to estimate $X$-ratio ($X_H$) within a crater 10s of km across. Dashed lines correspond to contours at the elevations of geologic features. Area within contours gives lake area estimate. Dividing topographic catchment area by lake area estimate gives $X_H + 1$.

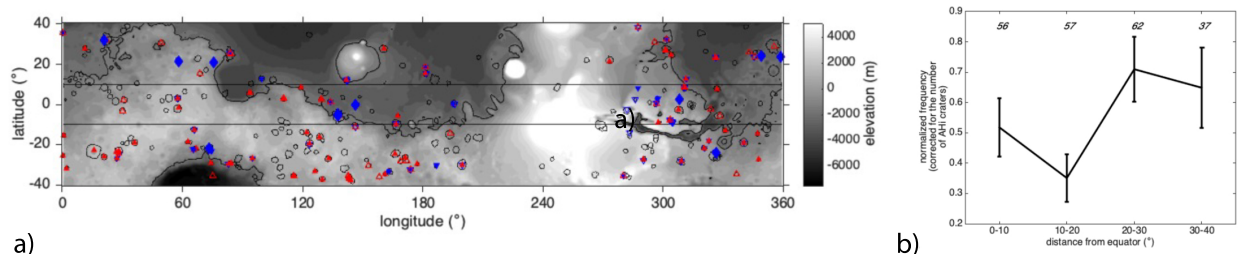

**Fig. 2. (a)** Aridity survey map (see Fig. S5 for details). Black contours locate young impact craters ("AHi" units from Tanaka et al. 2014) inspected for paleohydrology constraints. Black lines: -1500m elevation contour and ±10° latitude lines. Blue triangles = flat crater-bottom deposits (interpreted as lake deposits) (unfilled = candidate), blue diamonds = deltas/shorelines, blue circles= internal spillways, red filled triangles= alluvial fan toes, red unfilled triangles= channel-stops. **(b)** Distribution of frequency of craters with paleohydrologic evidence with distance from the equator (see Fig. S6 for details). Vertical bars correspond to $\sqrt{N}$ uncertainty. Numbers: per-bin sample size.

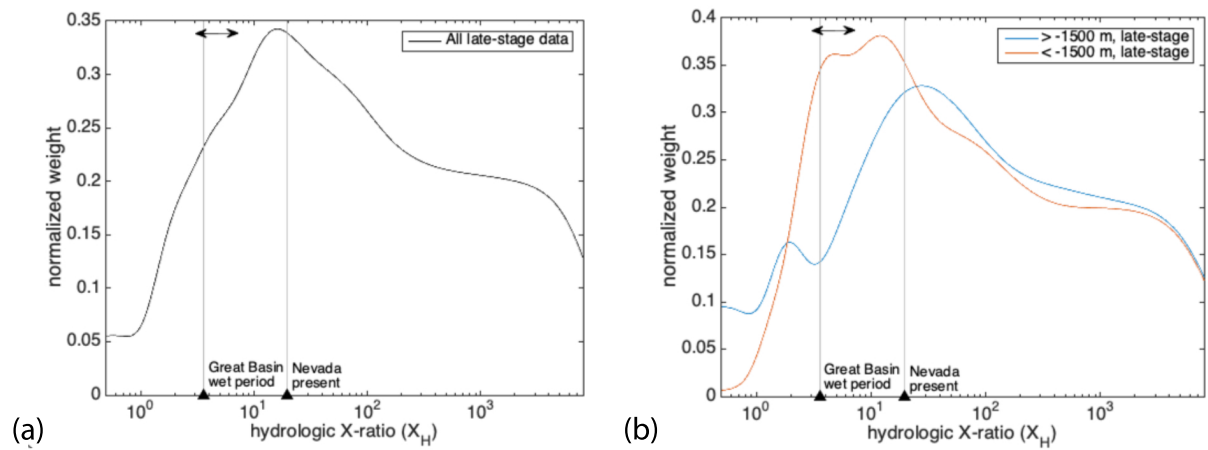

**Fig. 3.** Kernel-density estimates of late-stage aridity. The aridity that we estimated from late-stage geologic proxies is overall greater than that for early-stage rivers: double-headed arrows at top of plot correspond to Matsubara et al. 2011 estimate for early-stage rivers, $X_{H,ancient} = 5\pm2$. **(a)** Overall aridity increases with time. **(b)** High elevations dried out sooner. Here, paleo-aridity is assumed to simultaneously satisfy both geomorphic aridity upper limits and geomorphic aridity lower limits recorded within the same basin. Modern-Earth aridity values shown by black triangles are from Matsubara et al. (2011) (for Western Nevada). Log-uniform prior on $X_H = \{0.1, 10^4\}$; Fig. S7 shows sensitivity test.

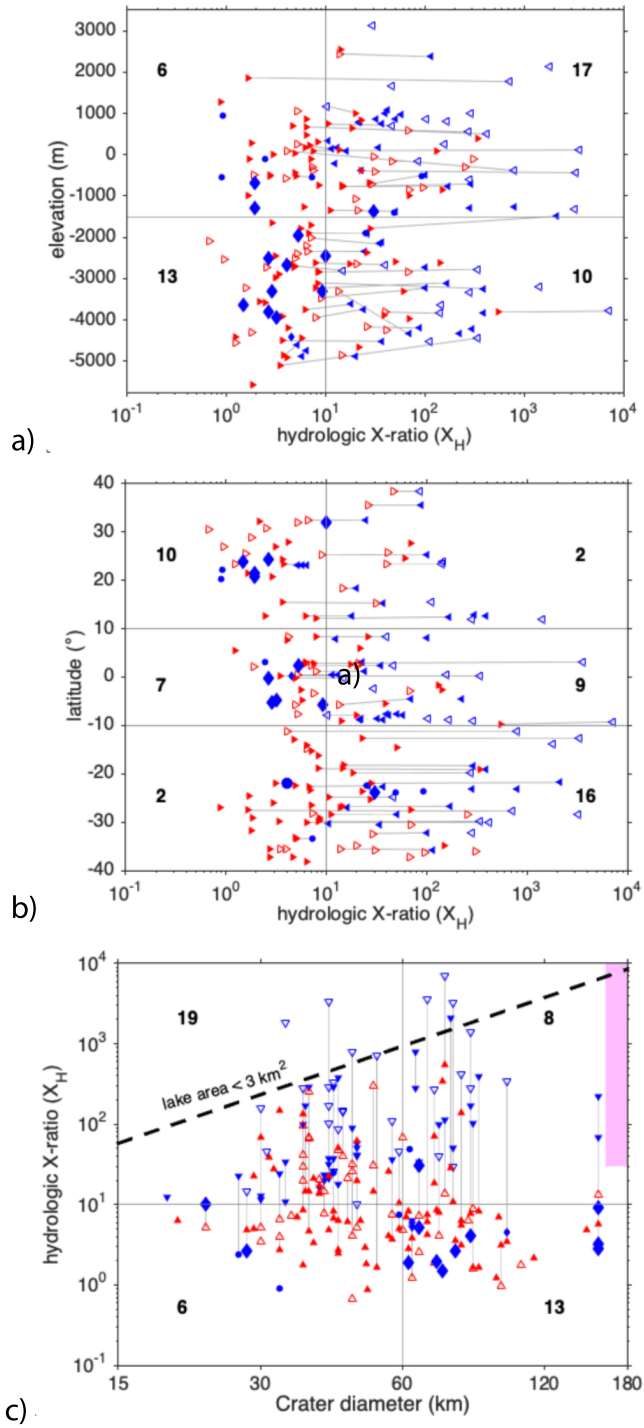

**Fig. 4.** Late-stage aridity constraints. Blue triangles are upper limits on aridity (e.g. lake deposit extent), red triangles are lower limits on aridity, (e.g. from alluvial fan termini), and blue diamonds are best-estimates of lake elevation (e.g., from a delta top). **(a)** $X_H$ versus elevation. Blue triangles=flat crater-bottom deposits (interpreted as lake deposits) (unfilled=candidate), blue diamonds=deltas/ shorelines, blue circles=internal spillways, red filled triangles=alluvial fan toes, and red unfilled triangles=channel-stops. Numbers correspond to the number of constraints lying entirely inside a rectangular region. Gray lines connect lowest and highest constraints for a single basin. (One -7500m data point is cropped). **(b)** $X_H$ versus latitude. **(c)** $X_H$ versus crater diameter. Black dashed line highlights where small lake deposits might be missed by survey.

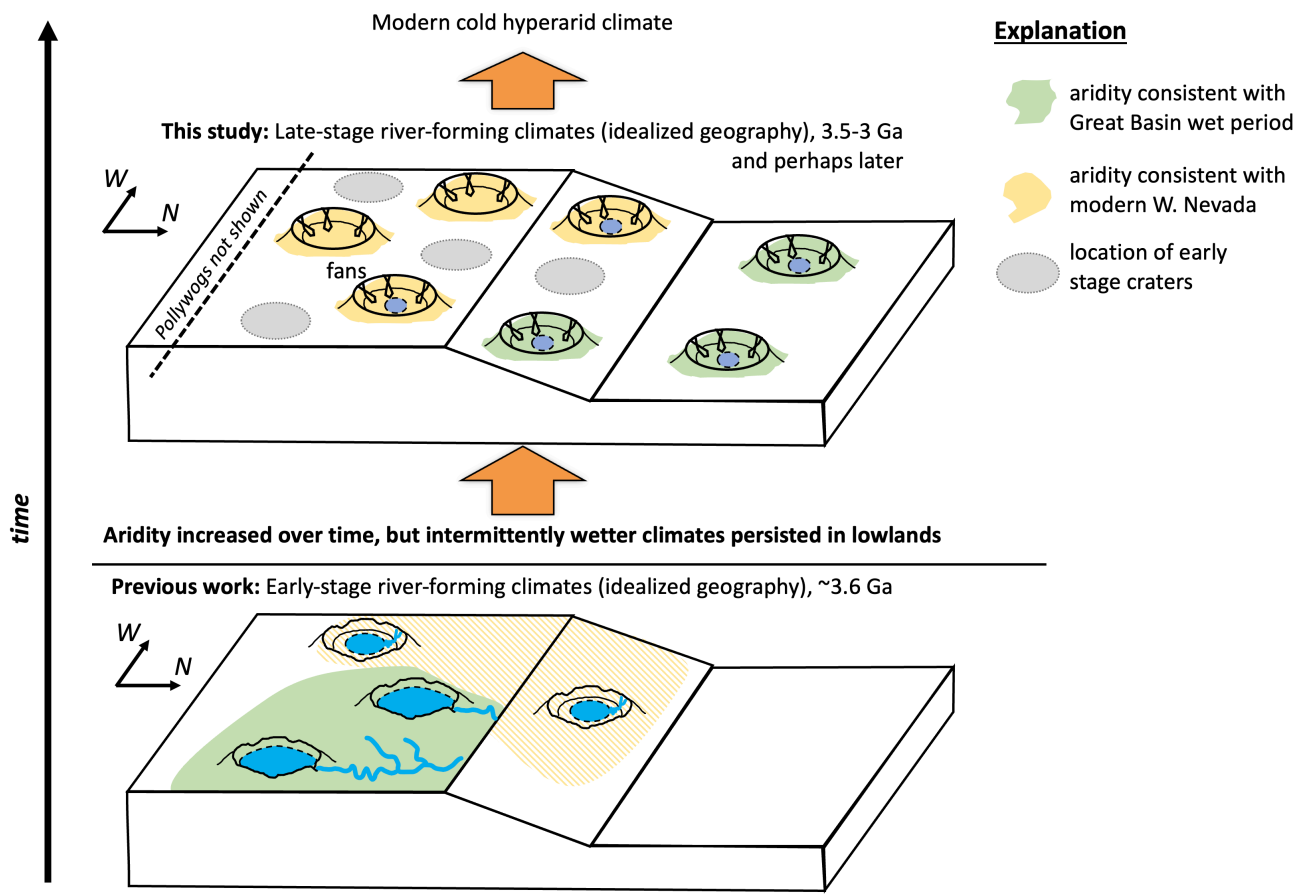

**Fig. 5.** Graphic summary of results.

Supporting Information for

**High and dry: billion-year trends in the aridity of river-forming climates on Mars**

Edwin S. Kite[1], Axel Noblet[2,3]

[1]University of Chicago, Chicago, IL, USA.

[2] Nantes Université, Nantes, France.

3. University of Western Ontario, Inst. Earth & Space Exploration, Canada.

Corresponding author: Edwin Kite (kite@uchicago.edu)

**Contents**

    Text S1
    Figures S1 to S10
    Tables S1 to S6

**Introduction**

Text S1 includes extended description of data and full details of statistical models.

Figures S1 to S8 include HiRISE DTM images (Figures S1 to S4) and extended description of data (Figures S5 to S10).

Tables S1 to S6 include extended description of data.

**Text S1.**

Trend of $X_H$ with crater size disfavors brief wet events.

Bigger craters were not any drier than small craters (there is a trend to bigger craters being wetter, $p<10^{-4}$, with $X_H<10$ found mainly in large craters, $p=0.003$) (Fig. 4c). This argues against brief wet events, for the following reason. Let the energetic upper limit on evaporation $E$ be $E_{max}$. For a given lake, if we have lake area $A$, drainage area $D$, and a terrain model, we know the minimum total runoff production $P_{tot}$ in order to flood area $A$. If the wet event was very brief (short timescale τ), then $P_{tot}/τ \gg E_{max}$. But if $P_{tot}/τ \gg E_{max}$, then during a wet event, small craters would fill up more than big craters. So, brief wet climates predict higher $X_H$ for bigger craters ($X_H \propto R_{crater}^{2/3}$) because small sinks fill quicker than big sinks. That is in contradiction to the observations. For fiducial values of $E_{max} = 1$ m/yr (Irwin et al. 2015) and $P_{tot} = 30$ m, each wet event must have lasted at least decades. This disfavors the scenario in which late-stage river-forming climates were powered by the greenhouse forcing from a single volcanic eruption, or the energy of a distant impact.

An alternative way of reaching the same result is as follows. Approximate a lake with volume $V$ as an inverted cone, $V = ⅓ π R_{lake}^2 h$, where $h$ is lake depth. Suppose that crater floors have the same floor slope $s$, so $V = ⅓ s π R_{lake}^3$. Considering a range of craters with different sizes, for uniform runoff production $V \propto R_{crater}^2$. Thus $R_{lake} \propto R_{crater}^{2/3}$ → $(R_{lake}/R_{crater}) \propto R_{crater}^{-1/3}$. Since for a closed basin $X_H \propto (R_{lake}/R_{crater})^2$, $X_H \propto R_{crater}^{2/3}$. Contrary to this prediction, $X_H$ is negatively correlated with $R_{crater}$ (Fig. 4c). This disfavors the brief-wet-event hypothesis. However, brief impact-triggered runoff appears to have occurred at some locations, such as Mojave crater (Goddard et al. 2014). Future work might use variable timescales, and full CTX DTM terrain models, to evaluate allowed combinations of timescale and $X_H$ (Stucky de Quay et al. 2020).

Details of analysis.

To check if the $X_H$ trends could be a statistical artifact, we used two approaches. First, we counted the number of hard constraints falling into rectangular regions in Fig. 4 (bold numbers in Fig. 4). We define a hard constraint to exclude channel-stops, candidate lake deposits, and basins where data permit $X_H$ on both sides of $X_H = 10$. (This is conservative in that channel-stops are probably good paleohydrologic constraints). This leaves 46 data points. The data are not evenly distributed between the rectangular regions. To find the probability that the trends result from chance, we resampled-with-replacement from the hard-constraint occurrences. Resampling showed $p = 0.0017$ for the latitude trend (lower $X_H$ at latitudes S of 10°S), $p = 0.0034$ for the crater-diameter trend (lower $X_H$ for crater diameter > 60 km), and $p = 0.0134$ for the elevation trend (lower $X_H$ at elevations < -1500m). As an alternative approach to uncertainty quantification, we randomly sampled aridities from a log-uniform prior on $X_H = \{0.1, 10^4\}$, clipped on a per-basin basis to satisfy the geologic constraints. (We did not resample basin occurrence in this approach, only the uncertainty on $X_H$ within each basin). This approach uses all 223 measurements and 118 basins. For each bootstrapped ensemble of basin aridities, we calculated the number of bootstrapped data falling into the rectangular regions shown in Fig. 4. Then, as for the first approach, we assessed trend agreement. From $10^4$ bootstraps, we found that in all cases the latitudinal trend is recovered, in all cases the crater-diameter trend is recovered, and in 9422 cases the elevation trend is recovered. We did a sensitivity test using a different log-uniform prior, $X_H = \{0.1, 10^{14}\}$. The sensitivity test results were unchanged for the latitude trend and crater-diameter trend and increased to 9695/10000 for the elevation trend. We conclude that systematic errors are more important than random error in our analysis.

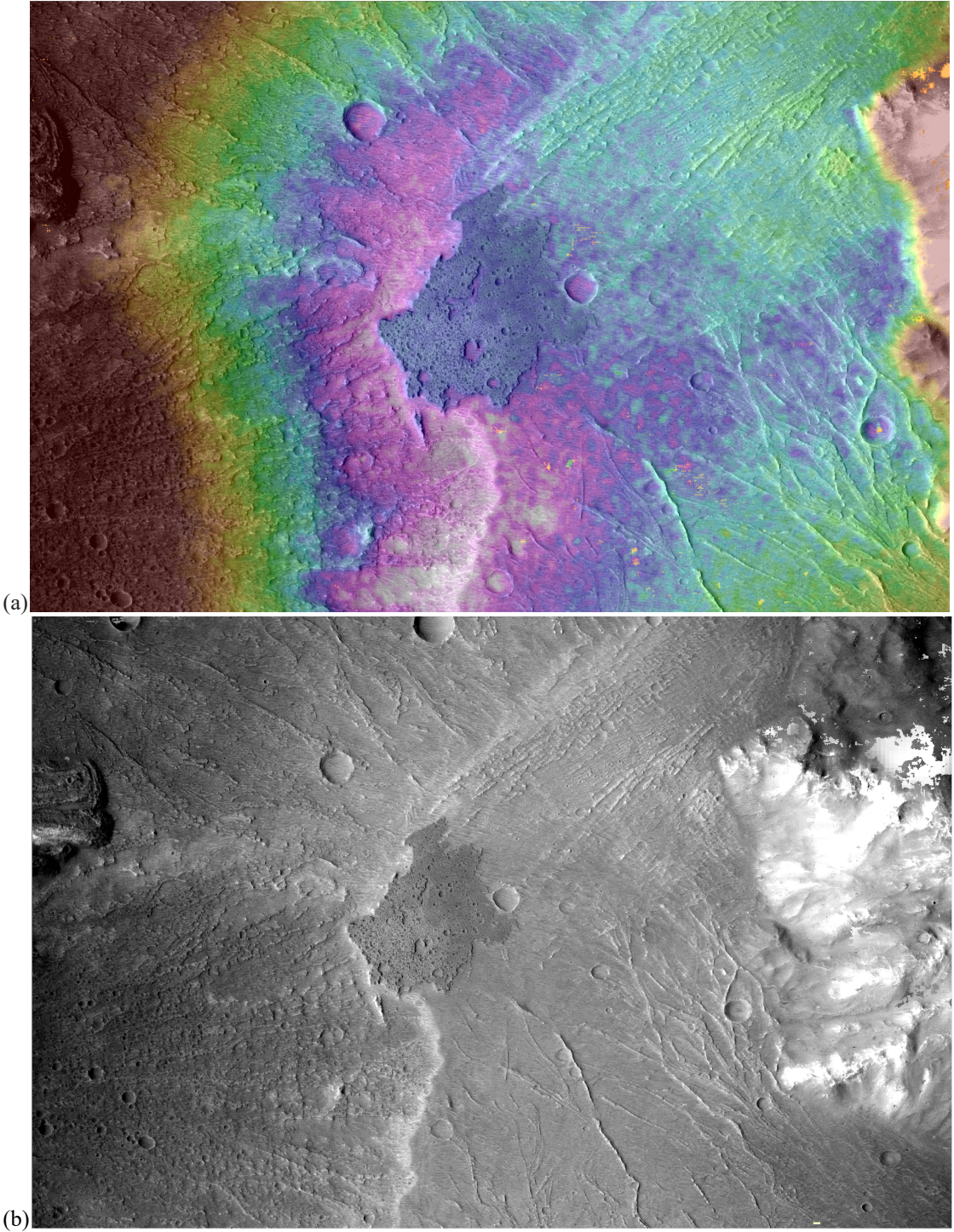

**Fig. S1. (a)** Example of flat crater-bottom deposit interpreted as a lake/playa deposit - mesa at terminus of wind-eroded alluvial fan deposits. 23°S 74°E. Image is 16.8 km across. Colors highlight elevation range between -1200m (red) and -1350m (white). CTX DTM (stereopair: F10_039889_1567_XN_23S286W and F12_040522_1566_XN_23S286W). **(b)** As (a), but without color elevation overlay.

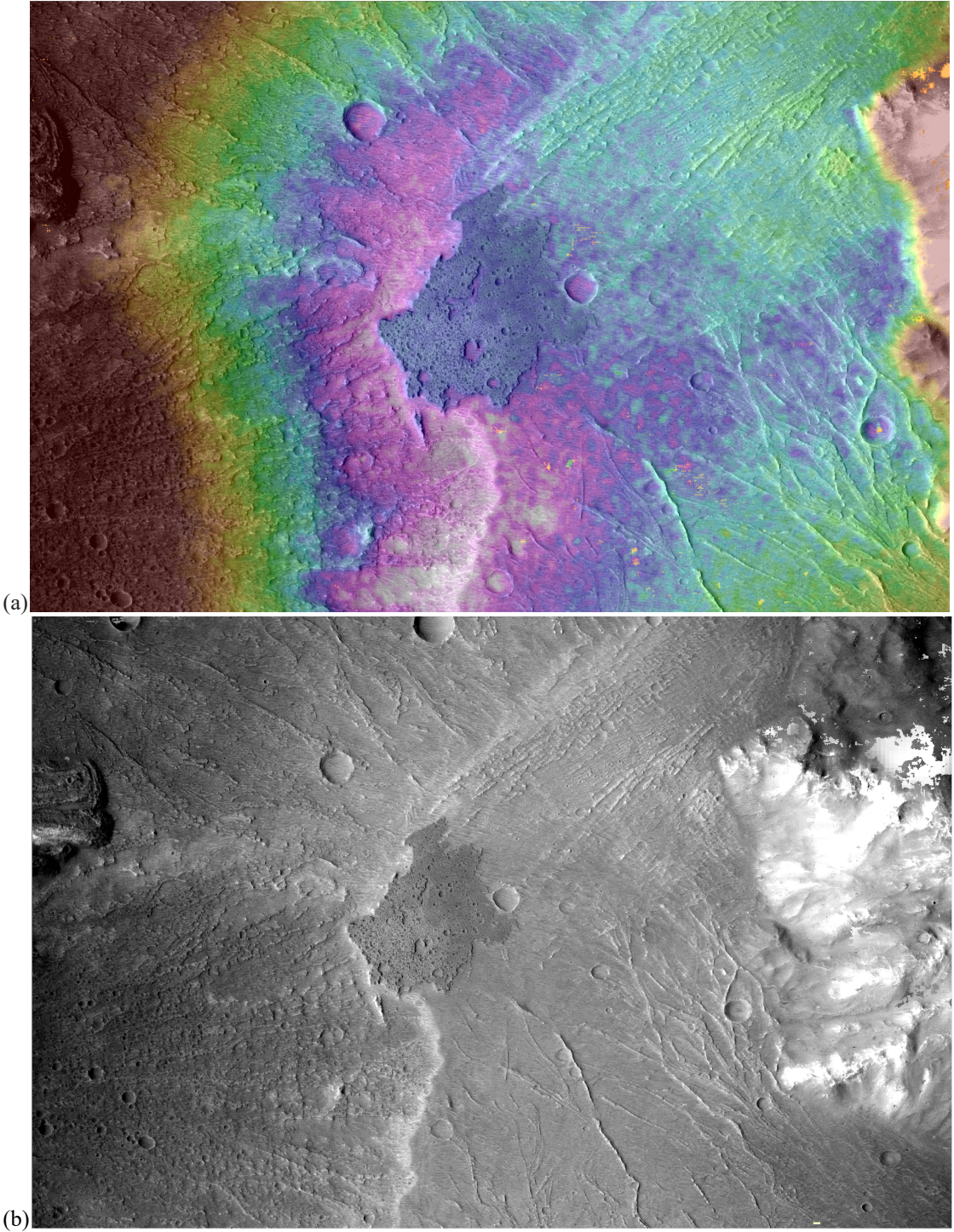

**Fig. S1. (a)** Example of flat crater-bottom deposit interpreted as a lake/playa deposit - mesa at terminus of wind-eroded alluvial fan deposits. 23°S 74°E. Image is 16.8 km across. Colors highlight elevation range between -1200m (red) and -1350m (white). CTX DTM (stereopair: F10_039889_1567_XN_23S286W and F12_040522_1566_XN_23S286W). **(b)** As (a), but without color elevation overlay.

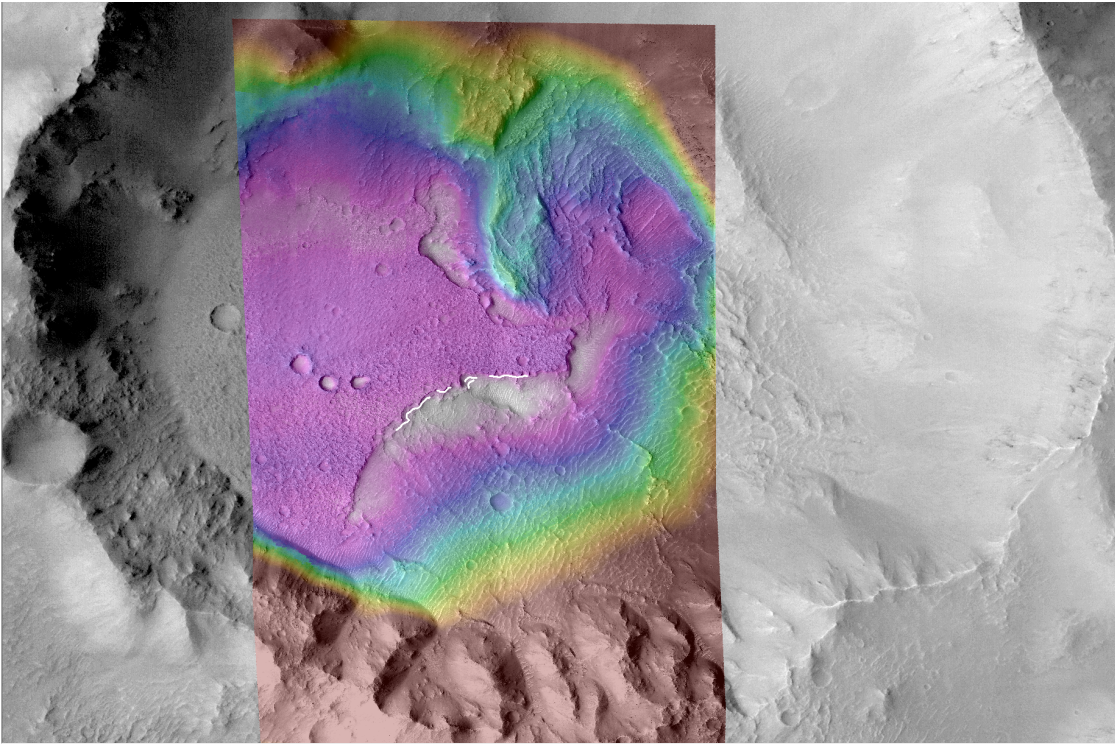

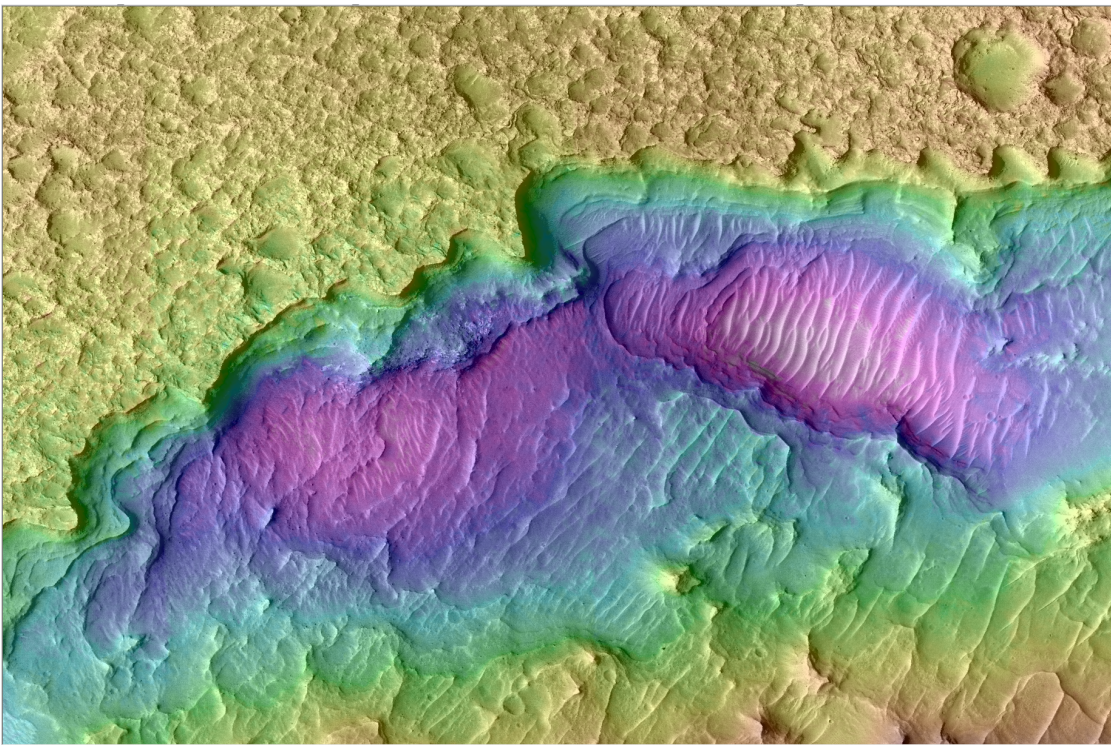

(c)

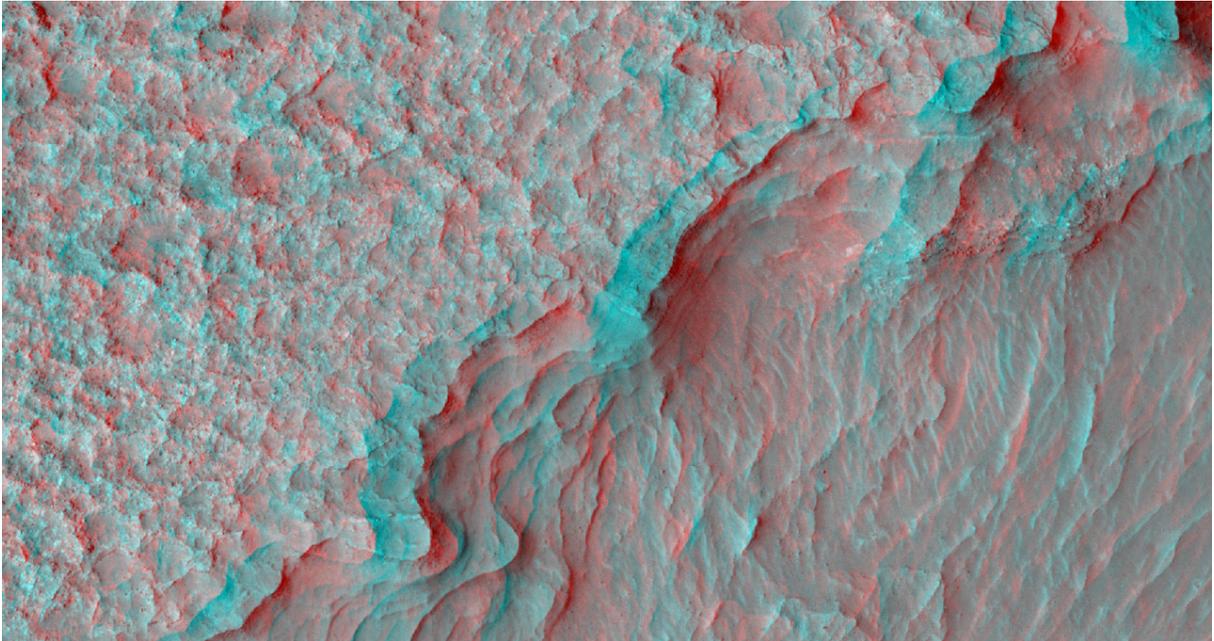

**Fig. S2.** Additional example of flat crater-bottom deposit interpreted as a playa/lake deposit (30°S 187°E). **(a)** Colored HiRISE DTM (ESP_065414_1495/ESP_065480_1495 stereopair) is 5.2 km across; grayscale background is CTX image. Colors highlight elevation range between 500 m (red) and 300m (white). The flat crater-bottom deposit is the purple mesa in the left center of the DTM. Note erosional alcoves in the S rim of the impact crater, and depositional ramp linking these alcoves to the flat-crater bottom deposit. The depositional ramp is topped by sinuous ridges, one of which feeds into the flat crater-bottom deposit. White lines trace layers whose elevation and orientation were measured (Fig. S8). **(b)** Close-up of the layered scarp corresponding to the white lines in panel (a). Image is 1.6 km across. Colors highlight elevation range between 348m (red) and 284m (white). **(c)** Red-blue anaglyph of the area in the left of panel (a). Image is 920m across. Nearly-horizontal mid-toned layers are exposed beneath the lighter-toned erosionally-resistant cap.

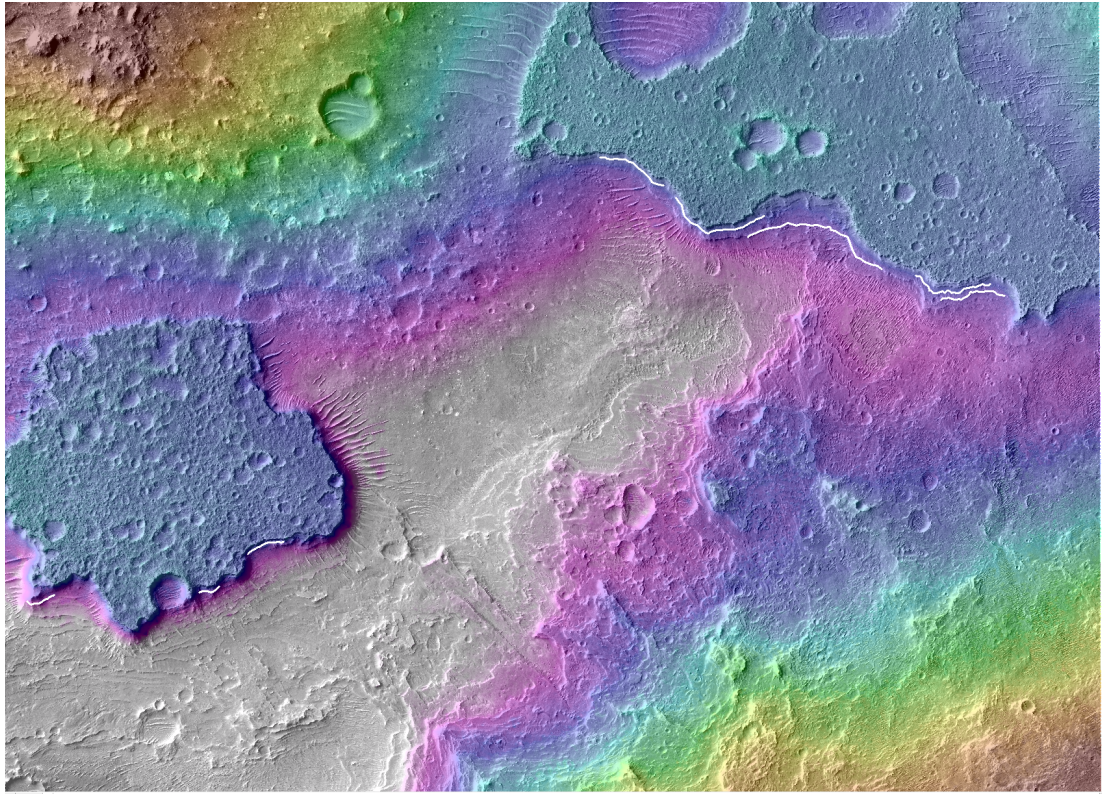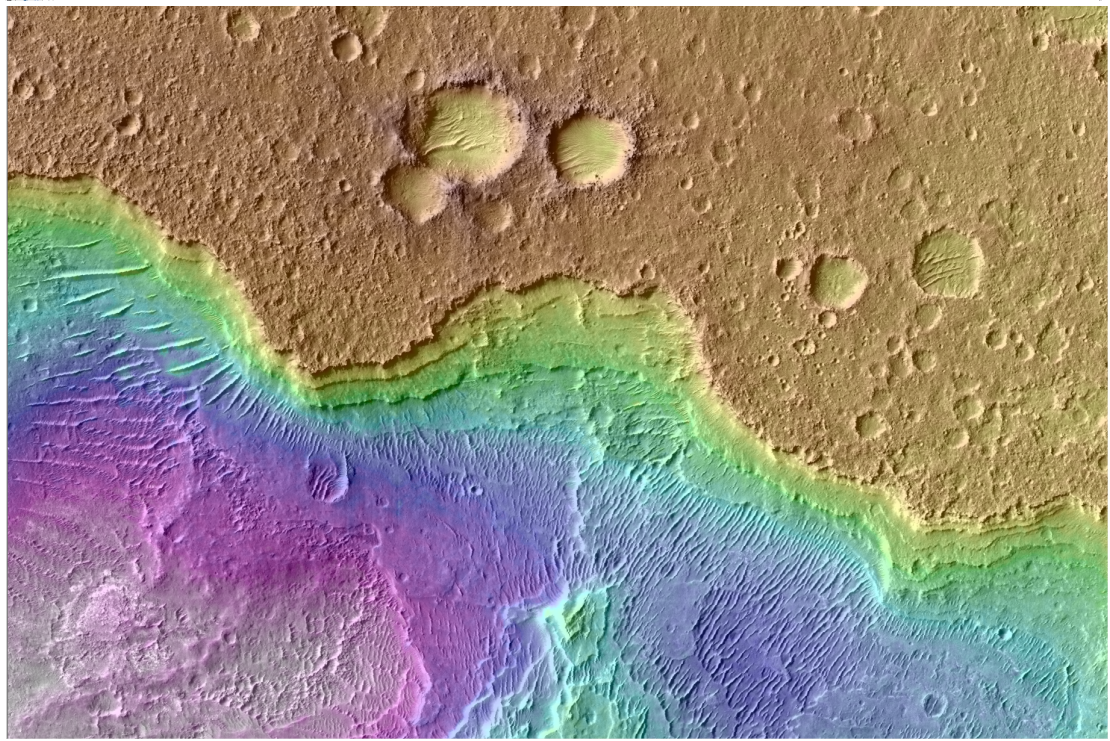

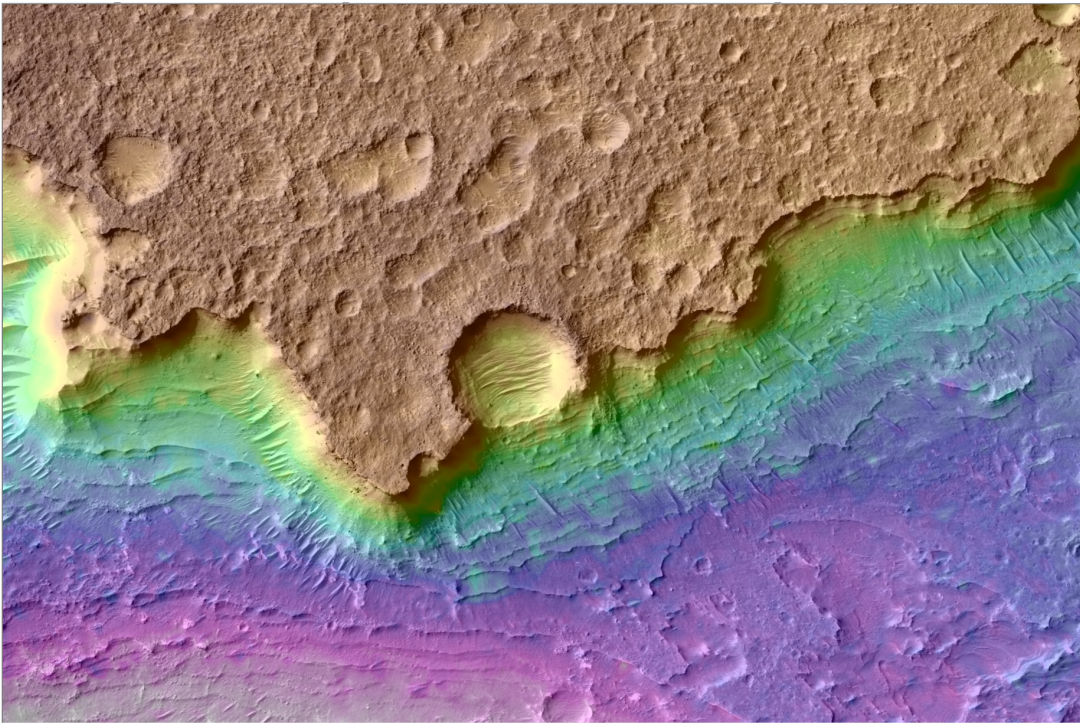

(c)

**Fig. S3.** Additional example of flat crater-bottom deposits interpreted as playa/lake deposits (18°S 323°E, Luba crater). **(a)** Image is 4.9 km across; colors highlight elevation range from -650m (red) to -750m (white) (ESP_072479_1615/ESP_072545_1615 stereopair). Note the alluvial fan deposit extending down from the bottom right. The white lines trace layers whose elevation and orientation were measured (Fig. S8). **(b)** Close-up of the area corresponding to the right cluster of white lines in panel (a). Image is 1.85 km across. Colors highlight elevation range between -705 m and -751 m. **(c)** Close-up of the area corresponding to the left cluster of white lines in panel (a). Image is 1.5 km across. Colors highlight elevation range between -707 m (red) to -779 m (white).

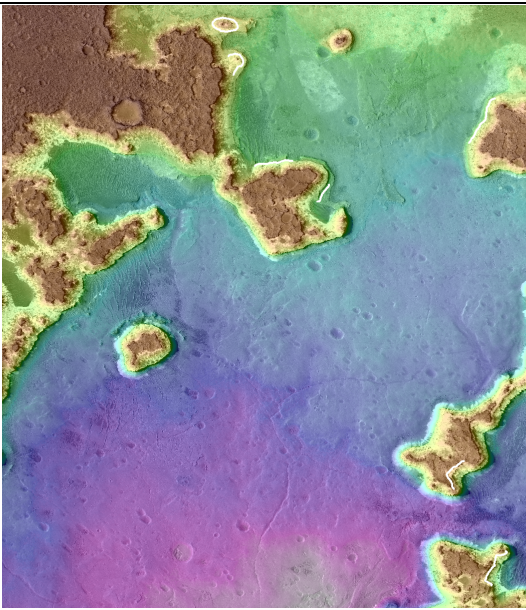

**Fig. S4.** Additional example of flat crater-bottom deposit interpreted as a playa/lake deposit (PSP_003526_1510/ PSP_003249_1510 stereopair , -29°S 309°E, Ritchey crater). Image is 5.9 km across. Colors highlight elevation range between -1355m (red) and -1551m (white). White lines indicate layer traces.

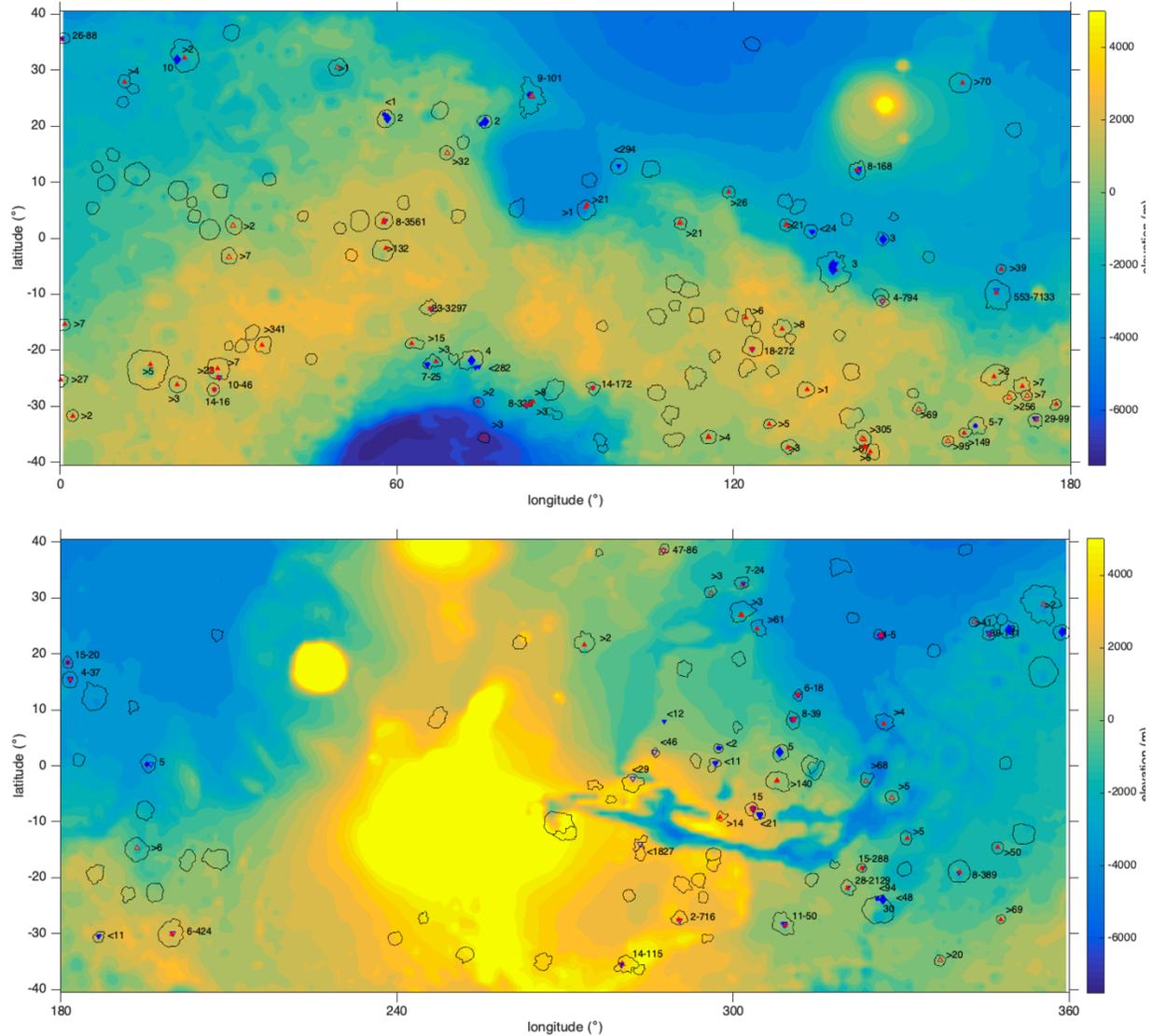

**Fig. S5.** Detailed version of Fig. 2a, showing $X_H$ constraints for different basins. Black contours locate young impact craters ("AHi" units from Tanaka et al. 2014) inspected for paleohydrology constraints. Blue triangles = flat crater-bottom deposits (interpreted as lake deposits) (unfilled = candidate), blue diamonds = deltas/shorelines, blue circles = internal spillways, red filled triangles = alluvial fan toes, red unfilled triangles = channel-stops. There is scatter in $X_H$ between nearby craters. Among other possibilities, the variability might be caused by lithological control (e.g., varying infiltration losses due to varying saturation hydraulic conductivity), the effect of crater shape on local meteorology (Steele et al. 2018), or feedback between erosion and snowmelt runoff (e.g., amplifying small differences in slope).

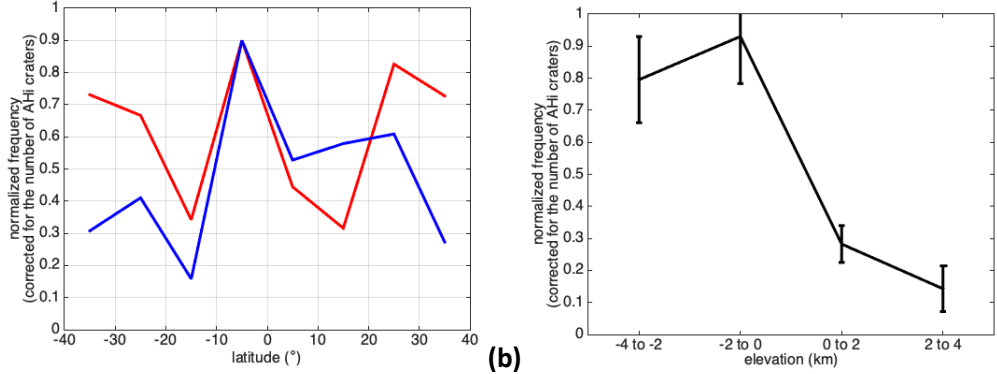

**Fig. S6. (a)** Detailed break-out of Fig. 2b, showing distribution with latitude of craters containing (blue line) geomorphic evidence for lower limits on lake extent and (red line) geomorphic evidence for upper limits on lake extent. **(b)** Distribution of frequency of craters with paleohydrologic evidence with elevation. Vertical bars correspond to $\sqrt{N}$ uncertainty. Numbers correspond to per-bin sample size.

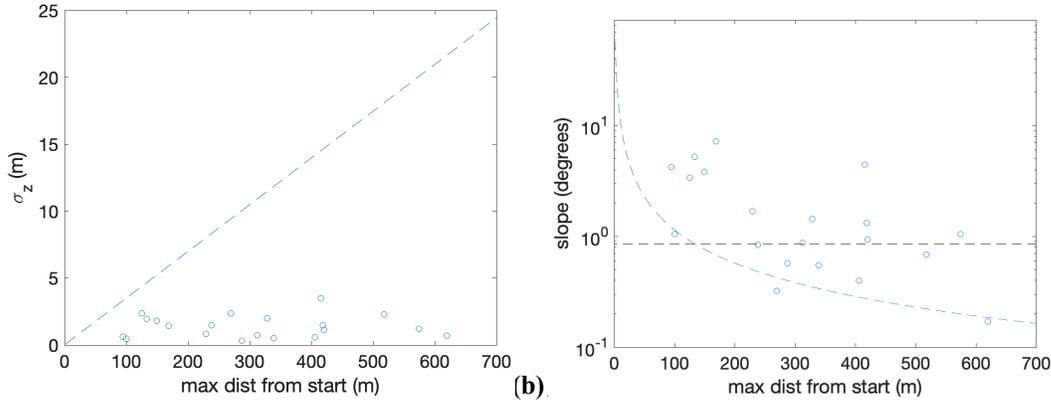

**Fig. S7.** Properties of layers within flat crater-bottom deposits interpreted as lake deposits. **(a)** The standard deviation of elevation of points along the trace is consistent with flat, taking into account tracing error and DTM uncertainty. **(b)** The median best-fit dip for traces >200 m long (black dashed line) is <1°, consistent with flat taking into account tracing error and DTM uncertainty. Orientations were calculated using the method of Lewis et al. (2008).

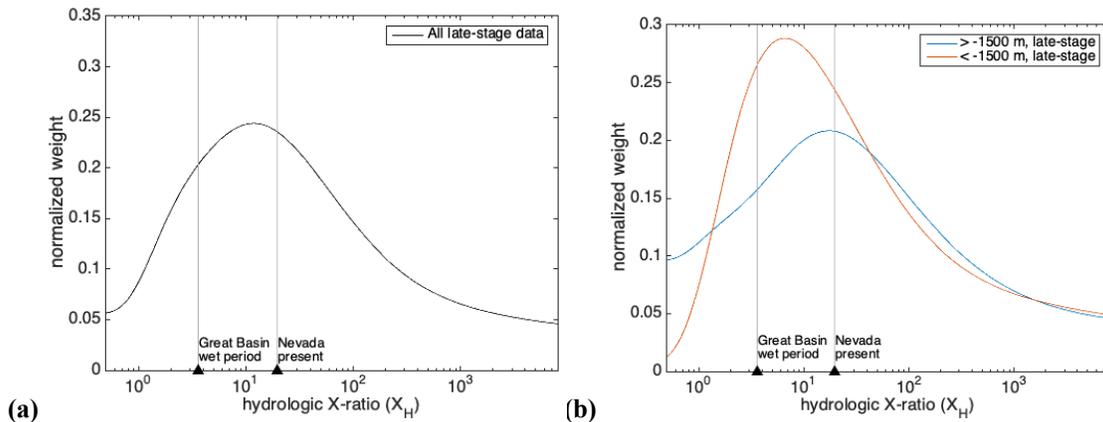

**Fig. S8.** Sensitivity test for Fig. 3, to show the effect of choosing a different log-uniform prior on $X_H$, specifically $\{0.1, 10^{14}\}$. Modern-Earth aridity values shown by black triangles are from Matsubara et al. (2011). **(a)** Overall late-stage aridity. **(b)** Kernel density estimate of late-stage $X_H$ changes with elevation.

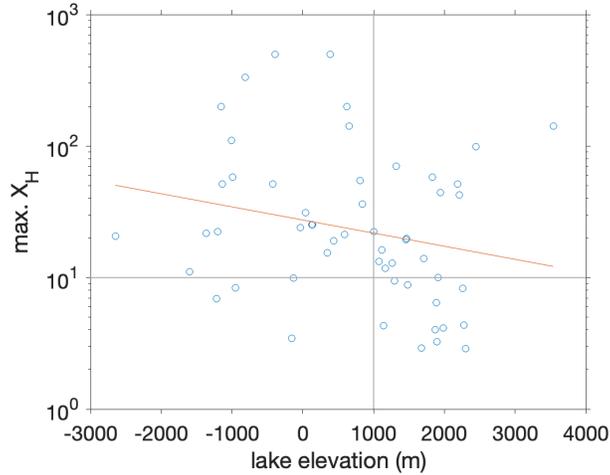

**Fig. S9.** Evidence for high-and-wet early stage dependence of aridity index on elevation. Upper limits on early-stage $X_H$ (high $X_H$ corresponds to greater aridity) from lake overspills using the data of Stucky de Quay et al. (2020). Paleolake locations from Table S1 in Stucky de Quay et al. (2020) Table S1 ($n = 54$) were interpolated in 8 pixels-per-degree MOLA data to obtain elevations. (Note that the larger and less selective dataset of Fassett & Head 2008 shows a weaker trend, with the same sign). The gray lines highlight the paucity of $X_H < 10$ constraints at elevations below +1000 m, and the red line corresponds to the least-squares best fit. The median $X_H$ lake-overspill bound above +1 km is $X_H < 12$, and the median $X_H$ lake-overspill bound above +1 km is $X_H < 25$.

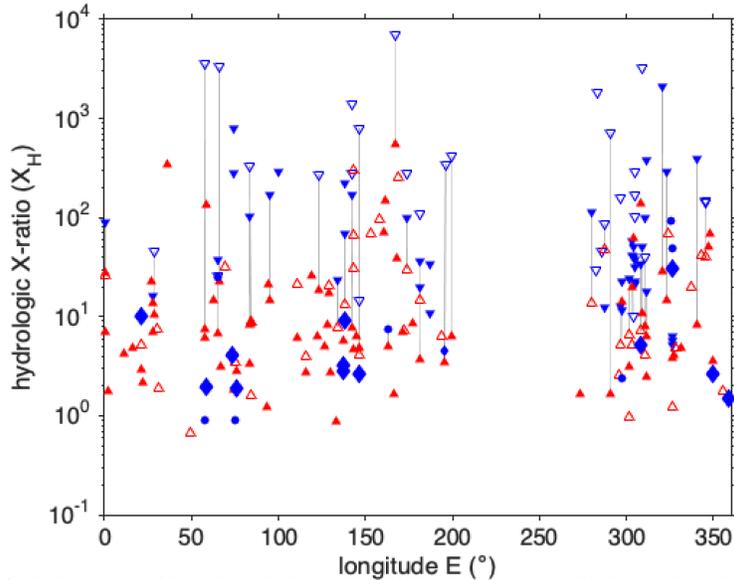

**Fig. S10.** Late-stage aridity constraints versus longitude (supplements Fig. 4). Blue triangles are upper limits on aridity (e.g. lake deposit extent), red triangles are lower limits on aridity, (e.g. from alluvial fan termini), and blue diamonds are best-estimates of lake elevation (e.g., from a delta top). Blue triangles=flat crater-bottom deposits (interpreted as lake deposits) (unfilled=candidate), blue diamonds = deltas/ shorelines, blue circles=internal spillways, red filled triangles=alluvial fan toes, and red unfilled triangles=channel-stops. Numbers correspond to the number of constraints lying entirely inside a rectangular region. Gray lines connect lowest and highest constraints for a single basin. The gap in longitude corresponds to the high, young volcanic province Tharsis.

| Location | Evidence (pre-fluviolacustrine-sed or syn-fluviolacustrine-sed impact craters internal to AHi impact rims) (ø = diameter) | Main-crater diam. (km) | Estimate [*] of Time Gap |
|---|---|---|---|
| 83°E 30°S (Nako) | Rivers/lakes activity postdate ejecta from large crater (40 km diameter?) to the E → long time gap | 43 | At least Gyr |
| 167°E 10°S (Reuyl) | ø = 6 km crater on SW side contains fan. | 86 | 0.4 Gyr |
| 303.5°E 7.6°S | ø = 2 km crater inside large crater (on W side) has inlet breach or alcove. | 45 | 0.2 Gyr |
| 311°E 8°N | Contains ø = 11 km crater that has alcoves, fans, and a possible lake deposit. [**] | 68 | 0.2 Gyr |
| 326°E 23°N (Wahoo) | Material with lineations perpendicular to high relief grades into wind-eroded material that is itself embayed by smoother, ramp material with much less wind erosion. Channel also postdates wind erosion. | 67 | At least Myr (to allow time for wind erosion) |
| 84°E 25.4°N (Peridier) | ø = 9 km crater on NW side is prefluvial. | 100 | 0.6 Gyr |
| 22°E 32°N (Cerulli) | Channel crosscuts ø = 7 km crater on the SW rim. | 130 | 0.2 Gyr |
| 144°E 38°S | Exit breach on ø = 2km crater on E side (internal to main crater). | 47 | 0.2 Gyr |
| 187°E 30°S | FCBD postdates ejecta from ø = 10 km crater on E rim (which itself contains an FCBD). | 31 | 6.5 Gyr (sic) |
| 297.5°E 3°N | Inlaid ø = 4 km crater inside W rim has exit breach into main crater. | 27 | 2 Gyr |
| 280°E 36°S | Probable exit-breach crater (ø = 6 km), inside N rim. | 69 | 0.6 Gyr |
| 326°E 26°S (Holden) | Noted by Irwin et al. (2015) and Kite et al. (2017). | 154 | At least Myr |
| Kite et al. 2017 sites (14 craters) | Interbedded craters. See Kite et al. 2017. | Varies between sites. | >(100–300) Myr |

**Table S1.** Evidence against a localized impact trigger for late-stage rivers and lakes. <u>Notes:</u> [*] Best-estimate time gap assuming modern impact flux (valid for the Amazonian, too low by a factor of 3.2 at 3.5 Ga) and the Hartmann chronology, using the nearest bin in the tables of Michael (2013). Assuming the count area for detection of interbedded craters is the entire crater (which will greatly understate the true time gap), and dividing results by a factor of 20 to take account of the fact that we only found interbedded craters in 12 of the ~219 craters that we surveyed. In reality synsedimentary and presedimentary impact craters are usually detected at/near the perimeter of sedimentary deposits, so the survey area is smaller than assumed here. Thus these estimates are crude and are likely biased low; even so, the timescales are long. [**] Additionally, PSP_008167_1885 shows a 700m-diameter crater prograded into by, and so predating, the Tyras fan. In addition to the craters tabulated here, sediments of uncertain origin (plausibly lacustrine) postdate a 1km-diameter crater within the SE of the 37km-diameter fan-bearing crater at 174°E 32°S.

| Constraint group long. (°) | Constraint group latitude (°) | Constraint group elevation (m) | Diameter of host crater (km) | Tightest upper bound on $X_H$ (0 = no constraint) | Tightest upper bound constraint type (see notes) | Tightest lower bound on $X_H$ | Tightest lower bound constraint type (see notes) |
|---:|---:|---:|---:|---:|---:|---:|---:|
| 65.41 | -22.57 | -1951.25 | 43 | 24.93 | 5 | 6.67 | 5 |
| 66.91 | -21.94 | -2450 | 80 | - | 0 | 3.10 | 1 |
| 73.33 | -21.86 | -2670 | 84 | 4.02 | 4 | 4.02 | 4 |
| 74.46 | -29.20 | -5570 | 50.4 | - | 0 | 1.82 | 1 |
| 94.87 | -26.71 | -775.5 | 37.4 | 171.75 | 5 | 14.47 | 5 |
| 133.04 | -27.08 | 1290 | 50.5 | - | 0 | 0.87 | 1 |
| 166.32 | -24.68 | -100 | 87 | - | 0 | 1.67 | 1 |
| 169.19 | -28.43 | -300 | 38 | - | 0 | 255.89 | 2 |
| 171.49 | -26.50 | -90 | 61 | - | 0 | 7.04 | 1 |
| 172.25 | -28.12 | -290 | 35 | - | 0 | 7.17 | 2 |
| 174.68 | -31.62 | -628.75 | 37 | 99.48 | 7 | 29.13 | 7 |
| 199.87 | -30.05 | 590 | 80 | 425.16 | 7 | 6.33 | 7 |
| 193.62 | -14.71 | -2190 | 83 | - | 0 | 6.42 | 2 |
| 146.49 | -11.26 | -145 | 47 | 794.46 | 7 | 4.07 | 7 |
| 123.30 | -19.78 | 590 | 70 | 272.83 | 7 | 18.25 | 7 |
| 122.12 | -14.13 | 850 | 20 | - | 0 | 6.40 | 1 |
| 66.01 | -12.62 | -412.5 | 42 | 3306.71 | 7 | 22.55 | 7 |
| 62.55 | -18.77 | -730 | 52 | - | 0 | 14.53 | 1 |
| 36.02 | -19.14 | 380 | 73 | - | 0 | 340.83 | 1 |
| 0.74 | -15.40 | -1700 | 36 | - | 0 | 7.00 | 1 |
| 347.19 | -14.59 | -2580 | 44 | - | 0 | 50.01 | 1 |
| 340.43 | -19.06 | -3170 | 87 | 388.70 | 5 | 8.31 | 5 |
| 331.12 | -12.94 | -2730 | 42 | - | 0 | 4.80 | 1 |
| 323.07 | -18.28 | -740 | 38 | 288.67 | 5 | 14.77 | 5 |
| 290.37 | -27.65 | 1820 | 53 | 717.47 | 7 | 1.66 | 7 |
| 309.27 | -28.45 | -1340 | 77 | 50.23 | 7 | 10.94 | 7 |
| 320.54 | -21.88 | -1635 | 76 | 2131.12 | 5 | 28.16 | 5 |
| 326.72 | -23.96 | -1375 | 65 | 30.10 | 3 | 30.10 | 3 |
| 347.84 | -27.47 | -1000 | 30 | - | 0 | 68.81 | 1 |
| 0.03 | -25.36 | -686 | 32 | - | 0 | 27.21 | 1 |
| 16.02 | -22.54 | -440 | 148 | - | 0 | 4.90 | 1 |
| 20.79 | -26.11 | 20 | 50 | - | 0 | 2.95 | 1 |
| 27.49 | -27.07 | 105 | 40 | 16.28 | 5 | 13.74 | 5 |
| 84.30 | -29.13 | -2830 | 71 | - | 0 | 8.46 | 1 |
| 83.06 | -29.80 | -2810 | 43 | 336.04 | 7 | 8.28 | 7 |
| 83.69 | -29.78 | -2890 | 39 | - | 0 | 3.35 | 1 |
| 30.03 | -3.37 | -366 | 57 | - | 0 | 7.48 | 2 |
| 58.07 | -1.75 | 85 | 37 | - | 0 | 132.18 | 1 |
| 137.81 | -5.02 | -3853.6 | 156 | 3.21 | 3 | 3.21 | 3 |
| 146.67 | -0.17 | -2663.3 | 28 | 2.67 | 3 | 2.67 | 3 |
| 166.77 | -9.52 | -3799.5 | 74 | 7136.97 | 7 | 553.09 | 7 |
| 167.73 | -5.60 | -3890 | 31.3 | - | 0 | 38.64 | 1 |
| 297.74 | -9.16 | 2560 | 40 | - | 0 | 13.97 | 1 |
| 303.54 | -7.73 | 1023.7 | 48 | 10.23 | 5 | 20.03 | 5 |
| 307.86 | -2.66 | -2210 | 80 | - | 0 | 140.26 | 1 |
| 323.77 | -2.75 | -2835 | 60 | - | 0 | 67.90 | 2 |
| 328.38 | -5.80 | -2500 | 60 | - | 0 | 4.85 | 2 |
| 30.80 | 2.19 | -500 | 70 | - | 0 | 1.88 | 2 |
| 57.71 | 3.06 | 131.3 | 68 | 3562.22 | 7 | 7.52 | 7 |
| 93.90 | 5.96 | -4650 | 39 | - | 0 | 21.36 | 1 |
| 93.60 | 5.55 | -4400 | 95 | - | 0 | 1.23 | 1 |
| 110.42 | 2.74 | -1307.5 | 46 | - | 0 | 21.19 | 2 |
| 119.13 | 8.29 | -2370 | 56 | - | 0 | 26.10 | 1 |
| 129.32 | 2.48 | -2648 | 40 | - | 0 | 20.63 | 2 |
| 195.54 | 0.26 | -4653.3 | 100 | 4.52 | 11 | 4.52 | 11 |
| 308.32 | 2.44 | -2043.3 | 65 | 5.24 | 3 | 5.24 | 3 |
| 310.75 | 8.31 | -2660.5 | 72 | 39.42 | 5 | 7.94 | 5 |
| 326.93 | 7.54 | -4910 | 56 | - | 0 | 4.04 | 1 |
| 68.89 | 15.20 | -417 | 47 | - | 0 | 31.53 | 2 |
| 142.24 | 12.12 | -3218.75 | 84 | 167.93 | 7 | 7.80 | 7 |
| 181.24 | 18.40 | -4875 | 41 | 19.99 | 5 | 14.52 | 5 |
| 181.63 | 15.47 | -4511 | 57 | 36.62 | 7 | 3.63 | 7 |
| 311.60 | 12.70 | -3692.25 | 44 | 17.66 | 5 | 6.22 | 5 |
| 11.41 | 27.93 | -4200 | 68 | - | 0 | 4.16 | 1 |
| 58.24 | 21.39 | -670 | 71 | 1.94 | 3 | 1.94 | 3 |
| 75.76 | 20.72 | -1475 | 62 | 1.90 | 3 | 1.90 | 3 |
| 83.84 | 25.38 | -3307 | 85 | 101.35 | 5 | 8.95 | 5 |
| 160.76 | 27.80 | -3970 | 70 | - | 0 | 69.77 | 1 |
| 301.48 | 26.95 | -2752.5 | 97 | - | 0 | 3.11 | 1 |
| 273.50 | 21.58 | -975 | 85 | - | 0 | 1.65 | 1 |
| 304.35 | 24.48 | -3300 | 48 | - | 0 | 61.37 | 1 |
| 326.34 | 23.25 | -4729.2 | 63 | 5.16 | 5 | 3.78 | 5 |
| 343.02 | 25.75 | -4240 | 37 | - | 0 | 41.02 | 2 |
| 345.73 | 23.60 | -3761 | 45 | 140.97 | 7 | 39.34 | 7 |
| 349.33 | 24.34 | -3850 | 78 | 2.66 | 3 | 2.66 | 3 |
| 355.44 | 28.84 | -4300 | 107 | - | 0 | 1.76 | 2 |
| 358.70 | 23.89 | -3650 | 73 | 1.47 | 3 | 1.47 | 3 |
| 0.38 | 35.60 | -4183 | 47 | 88.38 | 5 | 25.59 | 5 |
| 22.18 | 32.16 | -3550 | 114 | - | 0 | 2.13 | 1 |
| 20.77 | 31.91 | -2412.5 | 23 | 9.94 | 3 | 9.94 | 3 |
| 49.43 | 30.46 | -2100 | 47 | - | 0 | 0.67 | 2 |
| 287.58 | 38.46 | -170 | 44 | 86.21 | 7 | 46.78 | 7 |
| 296.09 | 30.84 | -2730 | 65 | - | 0 | 2.52 | 2 |
| 301.76 | 32.49 | -2342.5 | 33 | 24.40 | 5 | 6.53 | 5 |
| 2.28 | -31.73 | 282 | 37 | - | 0 | 1.78 | 1 |
| 75.40 | -35.66 | -7475 | 30 | - | 0 | 3.46 | 2 |
| 115.48 | -35.55 | -530 | 33 | - | 0 | 3.90 | 2 |
| 126.34 | -33.19 | 30 | 29 | - | 0 | 4.92 | 1 |
| 129.68 | -37.35 | -530 | 44 | - | 0 | 2.75 | 1 |
| 143.03 | -35.90 | -77.5 | 52 | - | 0 | 304.59 | 2 |
| 143.35 | -37.23 | 650 | 38 | - | 0 | 66.72 | 2 |

| | | | | | | | | |
|---|---|---|---|---|---|---|---|---|
| 144.37 | -38.10 | 470 | 49 | - | 0 | 6.30 | | 1 |
| 152.97 | -30.62 | -850 | 38 | - | 0 | 68.58 | | 2 |
| 158.23 | -36.31 | -780 | 37 | - | 0 | 94.75 | | 2 |
| 161.06 | -34.81 | -850 | 33 | - | 0 | 148.83 | | 1 |
| 163.10 | -33.45 | -530 | 59 | 7.35 | 10 | 4.95 | | 10 |
| 280.14 | -35.60 | 2407.5 | 74 | 114.71 | 5 | 13.59 | | 5 |
| 336.97 | -34.75 | -1083 | 37 | - | 0 | 19.79 | | 2 |
| 28.09 | -23.33 | 230 | 96 | - | 0 | 6.88 | | 1 |
| 26.97 | -23.63 | 830 | 29 | - | 0 | 22.55 | | 1 |
| 28.25 | -24.80 | 725 | 59 | 45.96 | 7 | 10.40 | | 7 |
| 128.55 | -16.16 | 315 | 62 | - | 0 | - | | 0 |
| 287.76 | 7.97 | -210 | 19 | 12.33 | 5 | - | | 0 |
| 74.29 | -23.06 | -1282.5 | 64 | 282.17 | 5 | - | | 0 |
| 304.79 | -8.76 | 838 | 42 | 21.42 | 7 | - | | 0 |
| 133.91 | 1.22 | -3740 | 41 | 23.76 | 5 | 7.80 | | 2 |
| 296.90 | 0.47 | 189 | 30 | 11.38 | 7 | - | | 0 |
| 297.52 | 3.19 | -240 | 27 | 2.42 | 5 | - | | 0 |
| 99.58 | 12.87 | -4230 | 43 | 294.41 | 5 | - | | 0 |
| 186.75 | -30.44 | 265 | 34 | 10.62 | 5 | - | | 0 |
| 325.83 | -23.71 | -520 | -1 | 93.63 | 10 | - | | 0 |
| 326.58 | -23.92 | -1400 | 62.2 | 48.48 | 10 | - | | 0 |
| 57.75 | 22.19 | 945 | 33 | 0.91 | 10 | - | | 0 |
| 75.07 | 20.36 | -540 | 33 | 0.10 | 10 | - | | 0 |
| 283.42 | -13.94 | 2140 | 34 | 1827.06 | 7 | - | | 0 |
| 282.08 | -2.37 | 3140 | 77 | 29.25 | 7 | - | | 0 |
| 286.06 | 2.38 | 1670 | 31 | 46.11 | 7 | - | | 0 |

**Table S2. Constrained basins.** Explanation. A zero or dash corresponds to "no constraint." Constraint types: 1 = Fan terminus. 2 = Channel terminus. 3 = Delta break-in-slope elevation or stepped-delta top. 4 = Fan toe/playa contact. 5 = Lake deposit. 7 = Candidate lake deposit. 10 = Overspilled contour. 11 = Shoreline feature. Additional notes. At Gale crater, we adopted Palucis et al. (2016)'s interpretation that the Pancake Delta is post-Mt. Sharp, fed by a small enclosed-basin catchment. The elevation of candidate shoreline features inside Nicholson crater (Salese et al. 2019; ESP_059361_1795) were treated as a best estimate of past lake level (a blue diamond in Fig. 4). At Saheki crater (Morgan et al. 2014), we observed the transition between an alluvial fan and a playa deposit. The elevation contour corresponding to this transition corresponds to our best-estimate of paleolake extent (a blue diamond in Fig. 4). Holden is omitted as the drainage area at the time deltas formed (Grant et al. 2008) is not known. At Peridier, channels extend topographically below the flat crater-bottom deposits that we interpret as lake deposits, perhaps corresponding to a later wet episode.

| Lon (°) | Lat (°) | Elevation (m) | Enclosed area (km²) | Topographic catchment area (km²) | $X_H$ | Constraint type |
|---|---|---|---|---|---|---|
| 65.47 | -22.62 | -1900 | 189.41 | 1452.20 | 6.67 | 1 |
| 66.91 | -21.94 | -2450 | 1226.63 | 5026.55 | 3.10 | 1 |
| 73.33 | -21.86 | -2670 | 1104.68 | 5541.77 | 4.02 | 4 |
| 74.46 | -29.20 | -5570 | 707.92 | 1995.04 | 1.82 | 1 |
| 94.88 | -26.67 | -776 | 71.01 | 1098.58 | 14.47 | 1 |
| 133.04 | -27.08 | 1290 | 1073.02 | 2002.96 | 0.87 | 1 |
| 166.32 | -24.68 | -100 | 2222.33 | 5944.68 | 1.67 | 1 |
| 169.19 | -28.43 | -300 | 4.41 | 1134.11 | 255.89 | 2 |
| 171.49 | -26.50 | -90 | 363.41 | 2922.47 | 7.04 | 1 |
| 172.25 | -28.12 | -290 | 117.77 | 962.11 | 7.17 | 2 |
| 177.44 | -29.63 | -730 | 113.56 | 1075.21 | 8.47 | 1 |
| 199.86 | -30.06 | 680 | 225.76 | 1655.00 | 6.33 | 1 |
| 193.62 | -14.71 | -2190 | 729.60 | 5410.61 | 6.42 | 2 |
| 146.49 | -11.23 | 90 | 342.08 | 1734.94 | 4.07 | 2 |
| 123.29 | -19.74 | 630 | 199.90 | 3848.45 | 18.25 | 1 |
| 122.12 | -14.13 | 850 | 42.46 | 314.16 | 6.40 | 1 |
| 66.00 | -12.63 | -390 | 58.84 | 1385.44 | 22.55 | 1 |
| 62.55 | -18.77 | -730 | 136.73 | 2123.72 | 14.53 | 1 |
| 36.02 | -19.14 | 380 | 12.24 | 4185.39 | 340.83 | 1 |
| 0.74 | -15.40 | -1700 | 127.22 | 1017.88 | 7.00 | 1 |
| 347.19 | -14.59 | -2580 | 29.81 | 1520.53 | 50.01 | 1 |
| 340.52 | -18.98 | -3100 | 638.67 | 5944.68 | 8.31 | 1 |
| 331.12 | -12.94 | -2730 | 238.81 | 1385.44 | 4.80 | 1 |
| 323.02 | -18.26 | -770 | 71.94 | 1134.11 | 14.77 | 1 |
| 290.35 | -27.58 | 1870 | 830.28 | 2206.18 | 1.66 | 1 |
| 309.29 | -28.50 | -1340 | 389.86 | 4656.63 | 10.94 | 1 |
| 320.49 | -21.96 | -1780 | 155.59 | 4536.46 | 28.16 | 1 |
| 326.72 | -23.96 | -1375 | 672.69 | 20921.80 | 30.10 | 3 |
| 347.84 | -27.47 | -1000 | 10.13 | 706.86 | 68.81 | 1 |
| 0.03 | -25.36 | -686 | 28.51 | 804.25 | 27.21 | 1 |
| 16.02 | -22.54 | -440 | 1598.28 | 9424.00 | 4.90 | 1 |
| 20.79 | -26.11 | 20 | 497.15 | 1963.50 | 2.95 | 1 |
| 27.48 | -27.07 | 110 | 85.25 | 1256.64 | 13.74 | 1 |
| 84.30 | -29.13 | -2830 | 418.55 | 3959.19 | 8.46 | 1 |
| 82.97 | -29.75 | -2850 | 156.54 | 1452.20 | 8.28 | 1 |
| 83.69 | -29.78 | -2890 | 274.44 | 1194.59 | 3.35 | 1 |
| 30.03 | -3.37 | -366 | 300.74 | 2551.76 | 7.48 | 2 |
| 58.07 | -1.75 | 85 | 8.07 | 1075.21 | 132.18 | 1 |
| 137.30 | -4.69 | -4440 | 654.36 | 4392.00 | 5.71 | 1 |
| 138.12 | -5.73 | -3310 | 224.70 | 3231.34 | 13.38 | 2 |
| 137.76 | -5.73 | -3300 | 947.56 | 9631.14 | 9.16 | 3 |
| 137.60 | -5.27 | -3300 | 5627.36 | 21693.47 | 2.85 | 3 |
| 137.69 | -4.67 | -3950 | 3037.70 | 12794.05 | 3.21 | 3 |
| 146.68 | -0.18 | -2690 | 223.54 | 1305.00 | 4.84 | 1 |
| 146.62 | -0.16 | -2500 | 355.41 | 1305.00 | 2.67 | 3 |
| 166.76 | -9.76 | -3813 | 4.53 | 2510.00 | 553.09 | 1 |
| 167.73 | -5.60 | -3890 | 19.41 | 769.45 | 38.64 | 1 |
| 297.74 | -9.16 | 2560 | 83.97 | 1256.64 | 13.97 | 1 |
| 303.41 | -7.82 | 1010 | 20.90 | 439.55 | 20.03 | 1 |
| 303.54 | -7.57 | 1050 | 150.22 | 925.88 | 5.16 | 2 |
| 307.79 | -2.70 | -1800 | 774.34 | 5026.55 | 5.49 | 1 |
| 307.94 | -2.61 | -2620 | 7.79 | 1101.00 | 140.26 | 1 |
| 323.77 | -2.75 | -2835 | 41.04 | 2827.43 | 67.90 | 2 |
| 328.38 | -5.80 | -2500 | 483.61 | 2827.43 | 4.85 | 2 |
| 30.80 | 2.19 | -500 | 1334.60 | 3848.45 | 1.88 | 2 |
| 57.67 | 3.21 | 180 | 267.57 | 1906.00 | 6.12 | 1 |
| 57.56 | 2.83 | 100 | 33.93 | 289.00 | 7.52 | 1 |
| 93.90 | 5.96 | -4650 | 53.42 | 1194.59 | 21.36 | 1 |
| 93.60 | 5.55 | -4400 | 3175.08 | 7088.22 | 1.23 | 1 |
| 110.48 | 2.77 | -1340 | 32.13 | 713.00 | 21.19 | 2 |
| 110.37 | 2.71 | -1275 | 234.31 | 1661.90 | 6.09 | 1 |
| 119.13 | 8.29 | -2370 | 90.87 | 2463.01 | 26.10 | 1 |
| 129.30 | 2.56 | -2650 | 33.82 | 622.00 | 17.39 | 1 |
| 129.34 | 2.41 | -2646 | 23.07 | 499.00 | 20.63 | 2 |
| 195.32 | 0.21 | -5100 | 1766.00 | 7853.98 | 3.45 | 1 |
| 308.37 | 2.42 | -1950 | 624.38 | 3896.00 | 5.24 | 3 |
| 308.30 | 2.47 | -2020 | 479.48 | 3896.00 | 7.13 | 2 |
| 310.96 | 8.39 | -2622 | 31.54 | 282.00 | 7.94 | 1 |
| 310.63 | 8.28 | -2620 | 797.94 | 4071.50 | 4.10 | 2 |
| 326.93 | 7.54 | -4910 | 488.49 | 2463.01 | 4.04 | 1 |
| 68.89 | 15.20 | -417 | 53.34 | 1734.94 | 31.53 | 2 |
| 142.07 | 12.11 | -3230 | 474.84 | 4179.00 | 7.80 | 1 |
| 181.23 | 18.40 | -4865 | 85.08 | 1320.25 | 14.52 | 2 |
| 181.60 | 15.46 | -4500 | 551.11 | 2551.76 | 3.63 | 1 |
| 311.53 | 12.68 | -3750 | 132.19 | 955.00 | 6.22 | 1 |
| 311.59 | 12.66 | -3570 | 442.29 | 1520.53 | 2.44 | 1 |
| 11.41 | 27.93 | -4200 | 703.74 | 3631.68 | 4.16 | 1 |
| 58.24 | 21.39 | -670 | 1346.17 | 3959.19 | 1.94 | 3 |
| 75.79 | 20.76 | -1300 | 1382.25 | 4014.00 | 1.90 | 3 |
| 75.74 | 20.68 | -1650 | 1038.20 | 4014.00 | 2.87 | 1 |
| 83.93 | 25.50 | -3225 | 2207.22 | 5674.50 | 1.57 | 2 |
| 84.09 | 25.23 | -3461 | 570.09 | 5674.50 | 8.95 | 2 |
| 160.76 | 27.80 | -3970 | 54.38 | 3848.45 | 69.77 | 1 |
| 301.51 | 26.91 | -2975 | 1865.34 | 7672.00 | 3.11 | 1 |
| 301.45 | 27.00 | -2530 | 3929.92 | 7672.00 | 0.95 | 2 |
| 273.50 | 21.58 | -975 | 2137.98 | 5674.50 | 1.65 | 1 |
| 304.35 | 24.48 | -3300 | 29.01 | 1809.56 | 61.37 | 1 |
| 326.47 | 23.39 | -4860 | 407.79 | 1951.00 | 3.78 | 1 |
| 326.30 | 23.29 | -4561 | 1402.82 | 3117.25 | 1.22 | 2 |
| 343.02 | 25.75 | -4240 | 25.59 | 1075.21 | 41.02 | 2 |
| 345.77 | 23.51 | -3810 | 32.70 | 1319.00 | 39.34 | 2 |
| 349.33 | 24.33 | -3900 | 1035.21 | 4778.36 | 3.62 | 1 |
| 349.32 | 24.35 | -3800 | 1306.13 | 4778.36 | 2.66 | 3 |
| 355.44 | 28.84 | -4300 | 3257.23 | 8992.02 | 1.76 | 2 |
| 358.70 | 23.89 | -3650 | 1696.69 | 4185.39 | 1.47 | 3 |
| 0.36 | 35.60 | -4171 | 65.25 | 1734.94 | 25.59 | 2 |
| 22.18 | 32.16 | -3550 | 3257.60 | 10207.03 | 2.13 | 1 |
| 20.77 | 31.92 | -2385 | 67.93 | 415.48 | 5.12 | 2 |
| 20.76 | 31.91 | -2440 | 37.99 | 415.48 | 9.94 | 3 |
| 49.43 | 30.46 | -2100 | 1037.81 | 1734.94 | 0.67 | 2 |
| 287.58 | 38.46 | -170 | 31.83 | 1520.53 | 46.78 | 2 |
| 296.09 | 30.84 | -2730 | 943.50 | 3318.31 | 2.52 | 2 |
| 301.72 | 32.44 | -2345 | 113.62 | 855.30 | 6.53 | 2 |
| 2.28 | -31.73 | 282 | 387.18 | 1075.21 | 1.78 | 1 |
| 75.40 | -35.66 | -7475 | 158.49 | 706.86 | 3.46 | 2 |
| 115.48 | -35.56 | -500 | 229.59 | 855.30 | 2.73 | 1 |
| 115.48 | -35.54 | -560 | 174.69 | 855.30 | 3.90 | 2 |
| 126.34 | -33.19 | 30 | 111.58 | 660.52 | 4.92 | 1 |
| 129.68 | -37.35 | -530 | 404.98 | 1520.53 | 2.75 | 1 |
| 142.94 | -35.86 | -40 | 20.24 | 640.00 | 30.62 | 2 |
| 143.12 | -35.95 | -115 | 5.64 | 1725.00 | 304.59 | 2 |
| 143.34 | -37.23 | 700 | 200.06 | 1134.11 | 4.67 | 1 |
| 143.36 | -37.23 | 600 | 16.75 | 1134.11 | 66.72 | 2 |
| 144.37 | -38.10 | 470 | 258.36 | 1885.74 | 6.30 | 1 |
| 152.97 | -30.62 | -850 | 16.30 | 1134.11 | 68.58 | 2 |
| 158.23 | -36.31 | -780 | 11.23 | 1075.21 | 94.75 | 2 |
| 161.06 | -34.81 | -850 | 5.71 | 855.30 | 148.83 | 1 |
| 163.13 | -33.45 | -520 | 459.12 | 2733.97 | 4.95 | 1 |
| 173.81 | -32.39 | -700 | 35.68 | 1075.21 | 29.13 | 2 |
| 280.18 | -35.50 | 2445 | 294.69 | 4300.84 | 13.59 | 2 |
| 336.97 | -34.75 | -1083 | 51.73 | 1075.21 | 19.79 | 2 |
| 28.09 | -23.33 | 230 | 918.41 | 7238.23 | 6.88 | 1 |
| 26.97 | -23.63 | 830 | 28.04 | 660.52 | 22.55 | 1 |
| 28.24 | -24.79 | 740 | 239.75 | 2733.97 | 10.40 | 1 |
| 128.55 | -16.16 | 315 | 322.45 | 3019.07 | 8.36 | 1 |
| 134.00 | 1.08 | -3945 | 149.98 | 1320.25 | 7.80 | 2 |

**Table S3. Fan terminus or delta-top elevation contours.** Explanation: Constraint types: 1 = Fan terminus. 2 = Channel-stop. 3 = Delta break-in-slope or Stepped-delta top. 4 = Fan terminus/playa intersection.

| Lon (°) | Lat (°) | Elev (m) | Lake area at overspill/shoreline (km²) | Topog. catchment area (km²) | Hydrologic X-ratio, $X_H$ | Diam. of host crater (km) | Notes |
|---:|---:|---:|---:|---:|---:|---:|---|
| 325.83 | -23.71 | -520 | 164.23 | 15540.70 | 93.63 | n.a. | 1. |
| 326.58 | -23.92 | -1400 | 422.84 | 20921.80 | 48.48 | 62 | 2. |
| 195.28 | 0.34 | -4420 | 1423.02 | 7853.98 | 4.52 | 100 | 3. |
| 297.42 | 3.18 | -100 | 10.53 | 36.00 | 2.42 | 27 | 4. |
| 57.75 | 22.19 | 945 | 448.14 | 855.30 | 0.91 | 33 | 5. |
| 75.07 | 20.36 | -540 | 440.47 | 836.00 | 0.90 | 33 | SW of Hargraves. |
| 163.07 | -33.45 | -540 | 221.64 | 1849.70 | 7.35 | 59 | |
| 65.33 | -22.50 | -1885 | 19.36 | 502.00 | 24.93 | 43 | SW of Harris. |

**Table S4. Overspilled contours + shoreline feature.** Notes: 1. Overspilled contour corresponds to a sediment trap upstream of the Eberswalde Delta. Topographic catchment is within the ejecta blanket of Holden crater (151 km diameter). 2. Overspilled contour within Eberswalde crater (Irwin et al. 2015). 3. This corresponds to ridges interpreted as a shoreline feature at Nicholson crater (Salese et al. 2019). 4. A 5×4 km crater-in-crater with an exit breach. 5. Exit-breach 33 km diameter crater immediately adjacent to a 71 km crater.

| Lon (°) | Lat (°) | Elev (m) | Area (km²) | Topog. catchment area (km²) | $X_H$ upper bound | Diam. of host crater (km) | Is Candidate | Notes |
|---|---|---|---|---|---|---|---|---|
| 287.76 | 7.97 | -210.00 | 21.28 | 283.53 | 12.33 | 19 | 0 | Not an 'AHi' crater |
| 65.53 | -22.67 | -2130.00 | 38.24 | 1452.20 | 36.98 | 43 | 0 | |
| 74.03 | -23.07 | -1300.00 | 7.57 | 2144.23 | 282.17 | 64 | 0 | Not an 'AHi' crater |
| 74.56 | -23.05 | -1265.00 | 1.75 | 1384.92 | 790.20 | 64 | 0 | |
| 94.86 | -26.74 | -775.00 | 6.36 | 1098.58 | 171.75 | 37.4 | 0 | |
| 340.35 | -19.15 | -3240.00 | 15.25 | 5944.68 | 388.70 | 87 | 0 | |
| 323.12 | -18.29 | -710.00 | 3.92 | 1134.11 | 288.67 | 38 | 0 | Luba crater |
| 309.37 | -28.54 | -1370.00 | 90.90 | 4656.63 | 50.23 | 77 | 0 | Ritchey crater |
| 320.58 | -21.80 | -1490.00 | 2.13 | 4536.46 | 2131.12 | 76 | 0 | Roddy crater |
| 27.49 | -27.08 | 100.00 | 72.70 | 1256.64 | 16.28 | 40 | 0 | |
| 138.21 | -4.57 | -4335.00 | 16.05 | 3586.85 | 222.52 | 156 | 0 | Gale crater |
| 138.00 | -4.48 | -4340.00 | 13.89 | 964.80 | 68.48 | 156 | 0 | Gale crater |
| 303.41 | -7.82 | 980.00 | 7.48 | 439.55 | 57.74 | 48 | 0 | |
| 303.52 | -7.86 | 1080.00 | 6.01 | 257.85 | 41.88 | 48 | 0 | |
| 303.67 | -7.69 | 996.00 | 4.69 | 189.39 | 39.38 | 48 | 0 | |
| 303.60 | -7.56 | 870.00 | 14.19 | 730.72 | 50.48 | 48 | 0 | |
| 304.70 | -8.62 | 800.00 | 12.43 | 290.61 | 22.38 | 42 | 0 | Elorza crater |
| 304.68 | -8.77 | 765.00 | 4.06 | 153.62 | 36.87 | 42 | 0 | Elorza crater |
| 304.91 | -8.83 | 790.00 | 16.85 | 377.70 | 21.42 | 42 | 0 | Elorza crater |
| 304.91 | -8.69 | 830.00 | 0.88 | * | * | 42 | 0 | Elorza crater |
| 304.88 | -8.64 | 855.00 | 9.85 | 322.74 | 31.77 | 42 | 0 | Elorza crater |
| 134.04 | 1.10 | -4170.00 | 28.75 | * | * | 41 | 0 | |
| 133.86 | 1.29 | -3720.00 | 51.77 | * | * | 41 | 0 | |
| 133.91 | 1.22 | -3740.00 | 53.33 | 1320.25** | 23.76** | 41 | 0 | |
| 296.90 | 0.38 | 177.00 | 27.48 | 381.90 | 12.90 | 30 | 0 | |
| 296.97 | 0.54 | 140.00 | 25.62 | 317.30 | 11.38 | 30 | 0 | |
| 297.61 | 3.20 | -380.00 | 24.18 | 572.56 | 22.68 | 27 | 0 | |
| 308.31 | 2.44 | -2160.00 | 110.35 | 3896.00 | 34.31 | 65 | 0 | Carmichael crater |
| 310.54 | 8.31 | -2825.00 | 4.79 | * | * | 72 | 0 | |
| 310.68 | 8.26 | -2830.00 | 17.21 | * | * | 72 | 0 | |
| 310.49 | 8.28 | -2770.00 | 35.28 | * | * | 72 | 0 | |
| 310.48 | 8.18 | -2740.00 | 40.46 | 4071.50** | 99.63** | 72 | 0 | |
| 99.58 | 12.87 | -4230.00 | 4.92 | 1452.20 | 294.41 | 43 | 0 | |
| 142.50 | 12.45 | -3100.00 | 8.01 | 1353.00 | 167.93 | 84 | 0 | Eddie crater |
| 181.25 | 18.40 | -4885.00 | 62.91 | 1320.25 | 19.99 | 41 | 0 | |
| 181.64 | 15.33 | -4515.00 | 67.83 | 2551.76 | 36.62 | 57 | 0 | |
| 311.76 | 12.61 | -3598.00 | 14.22 | * | * | 44 | 0 | |
| 311.72 | 12.73 | -3620.00 | 27.98 | 522.00** | 17.66** | 44 | 0 | |
| 311.52 | 12.71 | -3833.00 | 2.27 | * | * | 44 | 0 | |
| 311.54 | 12.73 | -3829.00 | 2.48 | 955.00** | 383.86** | 44 | 0 | |
| 83.74 | 25.85 | -3240.00 | 25.91 | * | * | 85 | 0 | Peridier crater |
| 83.58 | 25.65 | -3240.00 | 38.61 | * | * | 85 | 0 | Peridier crater |
| 83.53 | 25.52 | -3227.00 | 51.79 | * | * | 85 | 0 | Peridier crater |
| 83.50 | 25.41 | -3235.00 | 55.44 | 5674.50 | 101.35 | 85 | 0 | |
| 326.36 | 23.50 | -4930.00 | 218.31 | * | * | 63 | 0 | Wahoo crater |
| 326.54 | 23.28 | -4910.00 | 241.15 | * | * | 63 | 0 | Wahoo crater |
| 326.65 | 23.31 | -4890.00 | 283.23 | * | * | 63 | 0 | Wahoo crater |
| 326.61 | 23.21 | -4875.00 | 288.65 | 1951.00** | 5.76** | 63 | 0 | Wahoo crater |
| 326.02 | 23.22 | -4600.00 | 139.65 | 860.00 | 5.16 | 63 | 0 | Wahoo crater |
| 326.30 | 23.14 | -4750.00 | 32.75 | 239.00 | 6.30 | 63 | 0 | Wahoo crater |
| 0.41 | 35.61 | -4195.00 | 19.41 | 1734.94 | 88.38 | 47 | 0 | |
| 301.79 | 32.54 | -2340.00 | 33.68 | 855.30 | 24.40 | 33 | 0 | |
| 173.62 | -32.22 | -490.00 | 1.02 | 102.00 | 99.48 | 37 | 0 | |
| 186.68 | -30.53 | 185.00 | 26.03 | 907.92 | 33.88 | 34 | 0 | |
| 186.82 | -30.34 | 345.00 | 8.95 | 104.00 | 10.62 | 34 | 0 | |
| 280.09 | -35.46 | 2400.00 | 20.65 | * | * | 74 | 0 | Lampland crater |
| 280.10 | -35.71 | 2370.00 | 37.17 | 4300.84 | 114.71** | 74 | 0 | Lampland crater |
| 65.33 | -22.50 | -1890.00 | 18.51 | 502.00 | 26.13 | 43 | 1 | SW of Harris crater |
| 199.89 | -30.04 | 500.00 | 3.88 | 1655.00 | 425.16 | 80 | 1 | |
| 146.48 | -11.30 | -380.00 | 2.18 | 1734.94 | 794.46 | 47 | 1 | |
| 123.31 | -19.81 | 550.00 | 14.05 | 3848.45 | 272.83 | 70 | 1 | |
| 66.01 | -12.60 | -435.00 | 0.42 | 1385.44 | 3306.71 | 42 | 1 | |
| 283.42 | -13.94 | 2140.00 | 0.50 | 907.92 | 1827.06 | 34 | 1 | |
| 290.40 | -27.73 | 1770.00 | 3.07 | 2206.18 | 717.47 | 53 | 1 | Mazamba crater |
| 309.06 | -28.33 | -1270.00 | 1.15 | NaN | NaN | 77 | 1 | Ritchey crater |
| 309.14 | -28.32 | -1310.00 | 1.45 | 4656.63 | 3215.10 | 77 | 1 | Ritchey crater |
| 83.14 | -29.85 | -2770.00 | 4.31 | 1452.20 | 336.04 | 43 | 1 | Nako crater |
| 146.69 | -0.16 | -2800.00 | 82.62 | 1305.00 | 14.80 | 28 | 1 | Gunjur crater |
| 166.78 | -9.28 | -3786.00 | 0.30 | 2132.00 | 7136.97 | 74 | 1 | Reuyl crater |
| 282.08 | -2.37 | 3140.00 | 153.93 | 4656.63 | 29.25 | 77 | 1 | Perrotin crater |
| 303.65 | -7.79 | 1180.00 | 0.78 | 8.74 | 10.23 | 48 | 1 | Elorza crater |
| 304.78 | -8.60 | 850.00 | 3.13 | 323.00 | 102.23*** | 42 | 1 | Elorza crater |
| 304.71 | -8.88 | 799.00 | 1.30 | 218.00 | 166.58 | 42 | 1 | Elorza crater |
| 304.90 | -8.97 | 1010.00 | 1.31 | 378.00 | 288.30*** | 42 | 1 | Elorza crater |
| 57.91 | 3.15 | 114.00 | 0.53 | 1906.00 | 3562.22 | 68 | 1 | Leighton crater |
| 196.03 | 0.25 | -4440.00 | 5.14 | 1768.00 | 343.27 | 100 | 1 | Nicholson crater |
| 286.06 | 2.38 | 1670.00 | 16.02 | 754.77 | 46.11 | 31 | 1 | |
| 296.84 | 0.50 | 250.00 | 2.01 | 317.30 | 156.96 | 30 | 1 | |
| 310.94 | 8.38 | -2660.00 | 6.98 | 282.00 | 39.42 | 72 | 1 | |
| 142.06 | 11.99 | -3345.00 | 14.76 | 4179.00 | 282.09 | 84 | 1 | Eddie crater |
| 142.34 | 11.95 | -3200.00 | 2.96 | 4179.00 | 1411.42 | 84 | 1 | Eddie crater |
| 181.64 | 15.61 | -4520.00 | 22.90 | 2551.76 | 110.43*** | 57 | 1 | |
| 345.74 | 23.51 | -3839.00 | 9.29 | 1319.00 | 140.97 | 45 | 1 | |
| 345.68 | 23.78 | -3636.00 | 2.48 | 367.00 | 146.71 | 45 | 1 | |
| 287.59 | 38.46 | -170.00 | 17.43 | 1520.53 | 86.21 | 44 | 1 | Chukhung crater |
| 173.83 | -32.24 | -595.00 | 0.96 | 274.00 | 283.82 | 37 | 1 | |
| 28.26 | -24.82 | 710.00 | 58.22 | 2733.97 | 45.96 | 59 | 1 | Baum crater |

**Table S5. Flat Crater-Bottom Deposits Interpreted as Lake Deposits**. Notes: * When lake deposits were close to one another and appeared to be deposits from the same wet event, we combined their areas for the purpose of assessing lake size. In this table, deposit area is added to subsequent deposit area (with a common same drainage area) to give a combined hydrologic constraint. The area of the individual deposit is the difference between rows.  ** Combined lake areas for hydrology constraint. *** Combined with FCBDs for hydrology constraint.

| Name | Stereopair image 1 | Stereopair image 2 | DTM resolution |
|---|---|---|---|
| Luba1 | ESP_072479_1615 | ESP_072545_1615 | 2m |
| Luba2 | ESP_019467_1615 | ESP_018966_1615 | 2m |
| Ritchey | PSP_003249_1510 | PSP_003526_1510 | 1m |
| Unnamed_Magelhaens | ESP_065480_1496 | ESP_065414_1695 | 1m |

**Table S6. HiRISE DTMs made for this study.**